\documentclass{iopart}
 \ifx\pdfoutput\undefined
   \usepackage[dvips]{graphicx}
   \else
   \usepackage[pdftex]{graphicx}
   \fi
   \usepackage{epstopdf}
\usepackage{iopams}
\usepackage{color}
\definecolor{r}{rgb}{1,0,0}
\definecolor{b}{rgb}{0,0,1}

\begin{document}
\DeclareGraphicsExtensions{.pdf,.gif,.jpg}


\title{Scaling of $\mathcal{PT}$-asymmetries in viscous flow with $\mathcal{PT}$-symmetric  inflow and outflow}

\author{Huidan (Whitney) Yu$^{1,2}$, Xi Chen$^{1,3}$, Yousheng Xu$^4$, and Yogesh N. Joglekar$^5$}
\address{$^1$Department of Mechanical Engineering, Indiana University Purdue-University Indianapolis, IN 46202, USA}
\address{$^2$Richard G. Lugar Center for Renewable Energy, Indiana University Purdue-University Indianapolis, IN 46202, USA}
\address{$^3$Department of Physics, Zhejiang Normal University, Jinhua, China}
\address{$^4$School of Light Industry, Zhejiang University of Science and Technology, Hangzhou 310023, China}
\address{$^5$Department of Physics, Indiana University Purdue-University Indianapolis, IN 46202, USA}
\ead{$^1$whyu@iupui.edu,$^5$yojoglek@iupui.edu}

\begin{abstract}
 In recent years, open systems with balanced loss and gain, that are invariant under the combined parity and time-reversal ($\mathcal{PT}$) operations, have been studied via asymmetries of their solutions. They represent systems as diverse as coupled optical waveguides and electrical or mechanical oscillators. We numerically investigate the asymmetries of incompressible viscous flow in two and three dimensions with ``balanced'' inflow-outflow ($\mathcal{PT}$-symmetric) configurations. By introducing configuration-dependent classes of asymmetry functions in velocity, kinetic energy density, and vorticity fields, we find that the flow asymmetries exhibit power-law scaling with a single exponent in the laminar regime with the Reynolds number ranging over four decades. We show that such single-exponent scaling is expected for small Reynolds numbers, although its robustness at large values of Reynolds numbers is unexpected. Our results imply that $\mathcal{PT}$-symmetric inflow-outflow configurations provide a hitherto unexplored avenue to tune flow properties.
\end{abstract}


\section{Introduction}
\label{se:intro}
Open systems, where a system continuously exchanges information, such as energy and mass,  with its environment have been extensively studied due to their practical relevance and the theoretical interest they engender. Friction, Joule heating, and viscous drag~\cite{introphy} are ubiquitous examples of open-systems with losses, that exhibit one-way (energy) transfer to the environment. In contrast, open-systems with gains are rare, and are mostly realized in optical settings~\cite{lasers}. Over the past decade, theoretical research has predicted that open systems with ``balanced loss and gain'' exhibit novel properties that are absent in traditional open systems~\cite{bender1,review}. Such systems are described by equations of motion that are invariant under combined parity and time-reversal ($\mathcal{PT}$) operations; the resulting solutions, however, may or may not share that symmetry~\cite{review}. Although the field of $\mathcal{PT}$-symmetric quantum theories started out with spectral properties of non-Hermitian, $\mathcal{PT}$-symmetric continuum Hamiltonians, it has become evident that $\mathcal{PT}$ symmetric systems, classical or quantum, represents a special class of open systems that have both sources and sinks. Over the past three years, experiments on a wide variety of systems with balanced loss and gain - coupled optical systems~\cite{expt1,expt2,expt3,expt4}, coupled electrical oscillators~\cite{rlc}, and coupled mechanical oscillators~\cite{ajp} - have demonstrated the surprising  properties of such systems, such as unidirectional invisibility at optical frequencies~\cite{expt5}. This novel behavior arises from the {\it asymmetries} in the solutions of equations of motion.

Mathematically, symmetries of equations of motion, along with those of the boundary conditions, determine the symmetry properties of their solutions. For systems described by linear equations of motion, it is straightforward to obtain solutions with specific symmetries by linear superposition of linearly independent solutions. For example, the Schr\"{o}dinger equation for a quantum particle in an even potential $V(x)=V(-x)$ is invariant under the parity transformation $x\rightarrow -x$, and the corresponding eigenfunctions are either odd or even; if the initial state $\psi_0(x)$ of the particle has a definite parity symmetry, that symmetry is preserved during the time evolution~\cite{sakurai}. On the other hand, with an even initial wavefunction $\psi_0(x)=\psi_0(-x)$, if the potential $V(x)$ is not even, the time-evolved wave function $\psi(x,t)$ will develops an asymmetry $\rho(t)=\int dx |\psi(x,t)-\psi(-x,t)|$ that is determined by the asymmetry in the potential, $\rho_V=\int dx |V(x)-V(-x)|$. Thus, generically, if a  system with initial conditions that have a specific symmetry is evolved according to equations of motion that do not share the symmetry, the resulting solution will develop (time-dependent) asymmetries. The purpose of this paper is to investigate the dependence of such asymmetries in incompressible viscous flows subject to $\mathcal{PT}$-symmetric boundary conditions.

We emphasize here that experimentally investigated $\mathcal{PT}$-symmetric systems~\cite{expt1,expt2,expt3,expt4,expt5} had both dissipation and amplification of energy. Due to this balanced situation, they displayed a positive threshold for the loss/gain strength above which the solutions of equation of motion develop asymmetry. In contrast, the system considered in this paper is viscous and dissipative, with no attendant energy amplification, and therefore we expect that the asymmetries of flow solutions are always nonzero. The notions of inflow (mass-gain, energy-gain) and outflow (mass-loss, non-dissipative energy-loss) occur most naturally in fluid systems.  Traditional viscous flows are driven by upwind flow, pressure difference, or boundary movement, and therefore the steady-state velocity profiles at the inlet and the outlet are, in general, unrelated. In particular, flow symmetry properties in a system with specified inflow and outflow velocity profiles remain largely unexplored~\cite{caveat}. Viscous fluid flow with porous walls that act as inlets or outlets has been extensively studied ~\cite{berman,proudman,cox}, although not with symmetric boundary conditions that are investigated in this paper.

The incompressible fluid dynamics is governed by nonlinear Navier-Stokes (NS) equation,
\begin{equation}
\mathrm{Re} \partial_t{\bf u}+\mathrm{Re}({\bf u}\cdot\nabla){\bf u}=-\nabla p+\nabla^2{\bf u},
\label{eq:nse}
\end{equation}
and the continuity equation $\nabla\cdot{\bf u}=0$. Here the Reynolds number $\mathrm{Re}=u_p w/\nu$ is defined by the characteristic inlet velocity $u_p$, the width of the inlet $w$, and the kinematic viscosity $\nu$ of the fluid. For steady-state solutions, $\partial_t{\bf u}=0$, the effects of nonlinear, convective derivative are suppressed at small Reynolds number. At moderate to large values of Reynolds number, $\mathrm{Re}\gtrsim 1$, the nonlinear effects cannot be ignored and thus the symmetry properties of steady-state solutions of Eq.(\ref{eq:nse}) are not straightforward, and usually analytically intractable except in special cases~\cite{exactns1,exactns2}.

We use lattice Boltzmann method (LBM)~\cite{re:McNamara88,re:chen92,re:qian92} to numerically solve Eq. (\ref{eq:nse}) through an existing C program for a two-dimensional (2D) channel which had been previously validated. Originated from the lattice gas automata, the LBM has emerged as a popular alternative to model and simulate complex viscous flows ~\cite{rf:Chen98,rf:Aidun10}. The fundamental idea of this method is to construct simple kinetic models for spatially and temporally discretized particle-distribution functions that incorporate the essential physics of mesoscopic processes. The desired hydrodynamic variables in macroscopic equations are obtained from the moments of the particle-distribution functions~\cite{lbm1}. Although in the nearly-incompressible limit, the lattice Boltzmann equations recover the incompressible NS equations through the Chapmann-Enskog technique~\cite{rf:Chapmann70}, the computational philosophy of LBM is vastly different from traditional continuum NS-solvers. The main features that distinguish the LBM from continuum approaches are fourfold. i) The viscous diffusion in continuum solvers is replaced by a local relaxation process (collision operator) towards a local equilibrium state in LBM. ii) A linear convection operator in LBM generates the nonlinear macroscopic advection - the $({\bf u}\cdot\nabla){\bf u}$ term - through multiscale expansions. iii) For incompressible and isothermal single flows, such as the flow studied here, the particle distribution function is the only unknown to be determined, and the pressure distribution $p({\bf r})$ is obtained from the equation of state. iv) Computations required to obtain the particle distribution function are purely local. The distribution function at a point $({\bf r},t)$ depends only on its values at neighboring points, both spatially and temporally. Hence, the potential of LBM for parallelization is excellent. Another major advantage of the LBM is that its implementation is fairly simple and can be easily validated. Both two-dimensional (2D) and three-dimensional (3D) LBM codes used in this work were validated by testing their nearly identical agreement with the analytical solutions of Poiseuille flows.

In this paper, we introduce ``balanced inflow and outflow'' configurations of viscous flow in both 2D and 3D domains. The inlet and outlet velocity profiles, together with the geometry of the flow domain, are characterized by invariance under combined parity (reflection)  and time-reversal operations. We investigate the asymmetries of the resultant steady-state flow. A class of $\mathcal{PT}$-asymmetries in velocity, kinetic energy density, and vorticity is defined and its dependences on the Reynolds number and distinct configurations of ``balanced'' inflow and outflow are studied.

Our salient results are as follows: i) The asymmetries for three variables, the velocity ${\bf u}(\bf r)$, the kinetic energy density ${\cal E}({\bf r})$, and vorticity $\omega_z({\bf r})$, exhibit specific power-law scaling with the Reynolds number; the power-law exponent is determined by the asymmetry definition, but not the variable. ii) The asymmetries in the balanced inflow-outflow configuration are suppressed by orders of magnitude when compared with those in traditional fully-developed-flow configuration. iii) The power-law scaling is valid in both two and three dimensions. We emphasize that the total mass flux at the inlet is always equal to that at the outlet. Thus,  the phrase ``balanced inflow and outflow'' implies symmetry constraints on the velocity profile at the inlet and the outlet. 

The remainder of this paper is organized as follows. In Sec.~\ref{se:method} we present the formalism and the numerical method. We first define ``balanced inflow-outflow configurations'' and flow asymmetries, and then briefly describe how the LBM is used to solve NS equations on mesoscopic level. Section~\ref{se:results} presents results for 2D $\mathcal{PT}$-symmetric systems, including power-law scaling of asymmetries and steady-state velocity, vorticity, and kinetic energy density contours. Analytical considerations, discussed in Sec.~\ref{se:results}, imply that identical power-law scaling of asymmetries in all three variables is expected at small $\mathrm{Re}\lesssim 1$, but the scaling appears to hold at higher values of Reynolds numbers as well, $\mathrm{Re}\sim 10^2$. Section~\ref{se:3d} shows that the power-law scaling holds for 3D viscous flow in the laminar regime. We conclude the paper with a brief discussion in Sec.~\ref{se:disc}.


\section{Formalism and Numerical Method}
\label{se:method}

\subsection{Balanced inflow-outflow configuration and $\mathcal{PT}$ symmetry}
\label{subse:PT}
 We start with the definition of balanced inflow and outflow conditions and show how they relate to $\mathcal{PT}$ symmetries satisfied by the velocity profile at the boundaries. It is noted that for an incompressible flow in a rigid container, the mass inflow flux is equal to the outflow flux. Thus, we use the term balanced-inflow-outflow to denote a much stronger constraint on the fluid velocity at the inlet and outlet boundaries. As the first step, this subsection is confined to 2D geometries; corresponding formalism for the 3D case will be presented in Sec.~\ref{se:3d}.

\begin{figure}[htbp]
\begin{center}
\includegraphics[width=\columnwidth]{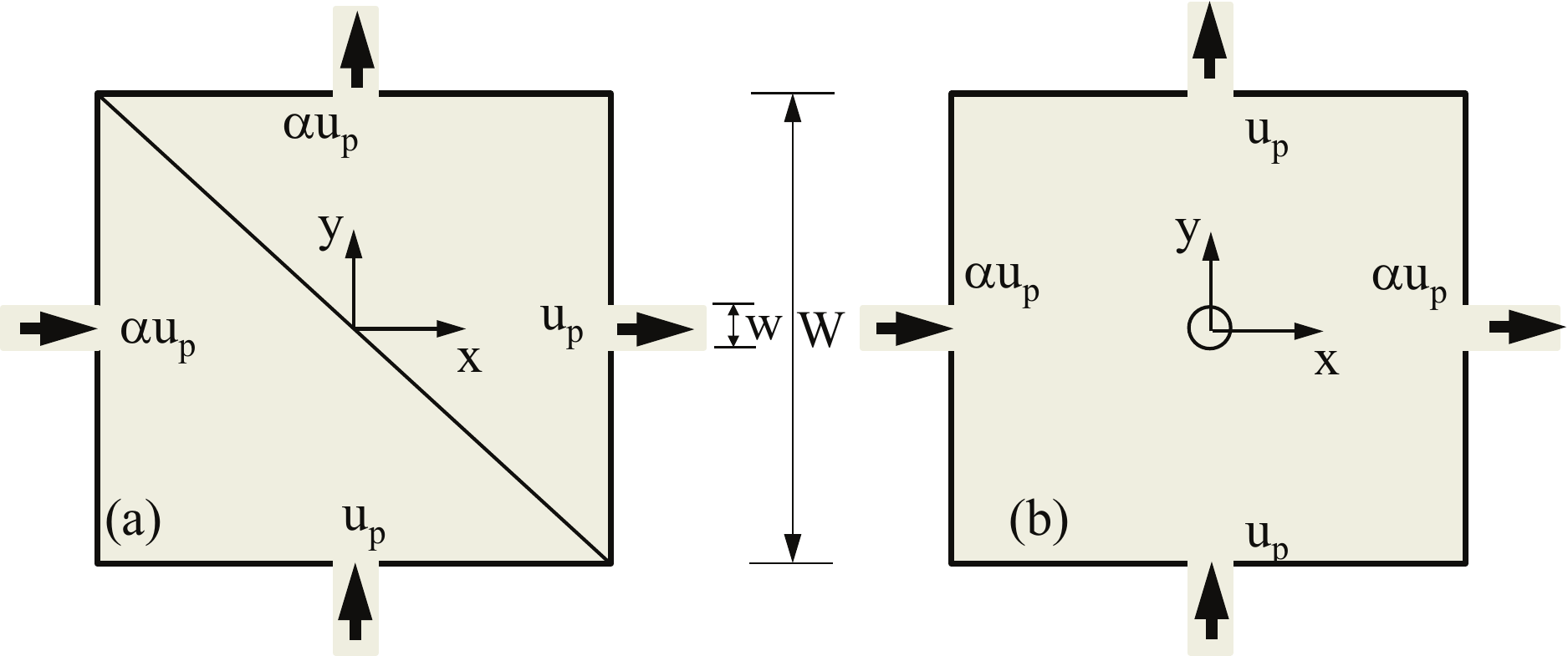}
\caption{Balanced inflow-outflow configurations possible in a square geometry. Panel (a) shows an {\it odd parity} and time-reversal symmetric configuration where parity $\mathcal{P}_O$ corresponds to reflection across the square diagonal line.  Panel (b) shows an {\it even parity} and time-reversal symmetric configuration where parity $\mathcal{P}_E$ corresponds to reflection through the origin. We impose identical velocity profile, i.e. uniform with speed or triangle/parabolic with maximum speed $u_p$ or $\alpha u_p$,  at each inlet and outlet although the results are independent of the profile.}
\label{fig:schematic}
\end{center}
\vspace{-5mm}
\end{figure}
We consider two square domains of width $W$ and inlet/outlet of width $w$. The origin of the coordinate system is located at the center of the flow domain, as shown in Fig.~\ref{fig:schematic}. Both panels in Fig.~\ref{fig:schematic} show balanced inflow-outflow configurations, but with different arrangements; here $0\leq\alpha\leq 1$ is an adjustable parameter. In panel (a), the two inflows, from the west and the south are balanced by two outflows to the north and the east respectively. Therefore, the configuration in panel (a) is symmetric about the northwest-southeast diagonal, shown by the diagonal line. Panel (b) shows another arrangement of balanced inflows and outflows from west to east and south to north respectively. This configuration is symmetric about the reflection through the origin, shown by the circle.

To quantify these symmetries, we define parity operators for both configurations. In panel (a), the inflow and outflow velocities ${\bf u}_b({\bf r})$ are invariant under the combined operation of reflection across the diagonal line, and time reversal. Therefore, an odd parity operator is introduced for the corresponding flow field
\begin{equation}
\label{eq:po}
\mathcal{P}_O:\left\{\begin{array}{c}
{\bf r}=(x,y)\rightarrow {\bf r}_O=(-y,-x)\\
{\bf v}({\bf r})=(u_x({\bf r}),u_y({\bf r}))\rightarrow (-u_y({\bf r}_O),-u_x({\bf r}_O))\\
\end{array}\right.
\end{equation}
The odd parity operator satisfies $\mathcal{P}_O^2=1$ and $\det\mathcal{P}_O=-1$. In contrast, the inflow and outflow velocities ${\bf u}_b({\bf r})$ in panel (b) are invariant under reflection through the origin (circle) and time reversal. Thus, an even parity operator is expressed for the corresponding flow field
\begin{equation}
\label{eq:pe}
\mathcal{P}_E:\left\{\begin{array}{c}
{\bf r}=(x,y)\rightarrow {\bf r}_E=(-x,-y)=-{\bf r}\\
{\bf u}({\bf r})=(u_x({\bf r}),u_y({\bf r}))\rightarrow -{\bf u}(-{\bf r})\\
\end{array}\right.
\end{equation}
The even parity operator satisfies $\mathcal{P}_E^2=1$ and $\det\mathcal{P}_E=+1$. In both cases, the velocity field is odd under time-reversal (or, more accurately, ``motion-reversal'') operation $\mathcal{T}{\bf u}({\bf r},t) = -{\bf u}({\bf r},t)$.  Under combined parity and time-reversal operations, the boundary velocity profiles satisfy $\mathcal{P}_O\mathcal{T}{\bf u}_b({\bf r})={\bf u}_b({\bf r}_O)$ in panel (a) and $\mathcal{P}_E\mathcal{T}{\bf u}_b({\bf r})={\bf u}_b({\bf r}_E)$ in panel (b). {\it This constraint, where the contain geometry is parity symmetric, and the boundary velocity profile is $\mathcal{PT}$-symmetric, defines a  ``balanced inflow-outflow configuration''}.

Note that $\alpha=0$ and $1/\alpha=0$ are two special cases where the number of inflow streams (and, equivalently outflow streams) reduces to one. In panel (a), the flow is driven from the south to the east if $\alpha=0$ or from the west to the north if $1/\alpha=0$. In panel (b), the flow is driven vertically from south to north, or horizontally from west to east. Apart from the inlets and outlets, fluid velocity at all other points on the boundary of the flow domain vanish, ${\bf u}_b=0$, due to the no-slip condition required by a viscous flow. When $\alpha=1$, the boundary velocity field is $\mathcal{PT}$-symmetric with respect to both parity operators.

\begin{figure}[htbp]
\centering
\includegraphics[width=0.95\columnwidth]{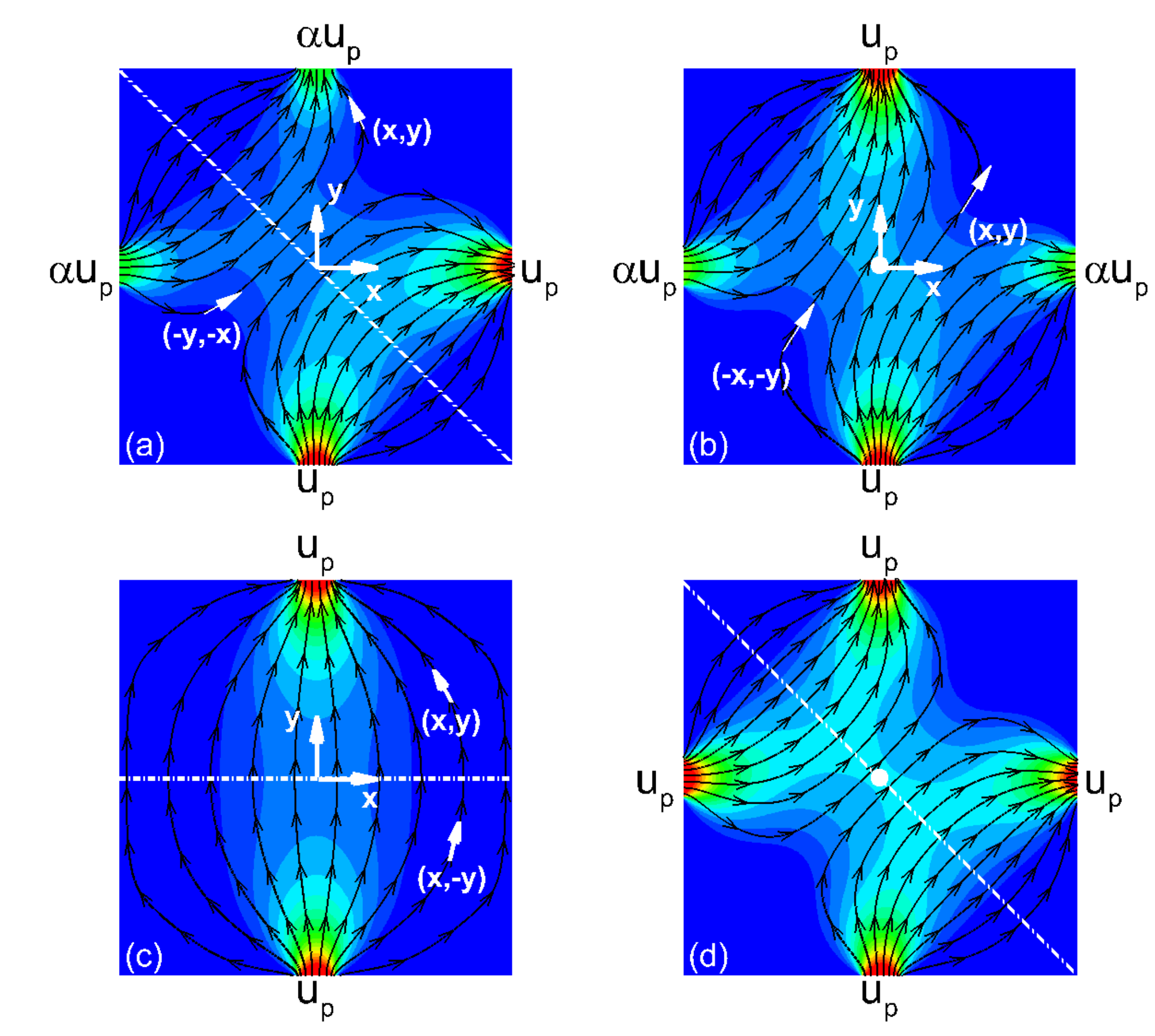}
\caption{Typical steady-state velocity fields ${\bf u}({\bf r})$ for different ``balanced inflow-outflow'' configurations with $w/W=0.1$ and a low Reynolds number ($10^{-2}$). Panel (a) shows velocity profile for odd-parity $\mathcal{PT}$-symmetric boundary condition, whereas panel (b) shows the velocity profile for even-parity $\mathcal{PT}$-symmetric boundary condition, both with $\alpha=0.58$; see Fig.~\ref{fig:schematic}. Panel (c) shows the flow lines for even-parity, $\alpha=0$ case, whereas panel (d) corresponds to the case $\alpha=1$. We note that in all cases, the steady-state velocity fields appear almost, but not exactly, $\mathcal{PT}$ symmetric.}
\label{fig:PTSymmetryTypes}
\end{figure}
To become more familiar with balanced inflow-outflow configurations, let us consider the resultant steady-state velocity field in the presence of $\mathcal{PT}$-symmetric boundary velocity profiles. (The details of numerical simulations are given in Sec.~\ref{subse:LBM}). Fig.~\ref{fig:PTSymmetryTypes} shows four typical steady-state velocity fields ${\bf u}({\bf r})$ in a square domain with $w/W=0.1$ at a low Reynolds number ($10^{-2}$). Panel (a) shows ${\bf u}({\bf r})$ for $\mathcal{P}_O\mathcal{T}$-symmetric boundary velocity profile, whereas panel (b) shows corresponding results for $\mathcal{P}_E\mathcal{T}$-symmetric configuration. These results are for $\alpha=0.58$. Panel (c) shows the results for $\mathcal{P}_E\mathcal{T}$-symmetric configuration with $\alpha=0$ and panel (d) has the velocity profile are for $\alpha=1$.  It is clear from Fig.~\ref{fig:PTSymmetryTypes} that starting from $\mathcal{PT}$-symmetric boundary conditions ${\bf u}_b({\bf r})$, the resultant steady-state velocity profile is close to, if not exactly, $\mathcal{PT}$ symmetric; recall that a velocity field ${\bf u}({\bf r},t)$ is $\mathcal{PT}$-symmetric if and only if it satisfies $\mathcal{PT}{\bf u}({\bf r},t)={\bf u}(\mathcal{P}{\bf r},t)$~\cite{caveat}. In the following subsection, the deviations from the $\mathcal{PT}$-symmetric profile for the velocity field, as well as for kinetic energy density and the vorticity fields are quantified.


\subsection{$\mathcal{PT}$ asymmetries of the viscous flow}
\label{subse:rho}
To quantify the deviation from $\mathcal{PT}$-symmetric field, we introduce a class of dimensionless $\mathcal{PT}$-{\it asymmetry} functions. The asymmetry of velocity field is characterized by $\Delta{\bf u}({\bf r})=\mathcal{PT}{\bf u}({\bf r})-{\bf u}(\mathcal{P}{\bf r})$. We define dimensionless $\mathcal{PT}$ {\it asymmetries} $\rho^u_O$ and $\rho^u_E$ in the steady-state velocity field ${\bf u}({\bf r})$ as
\begin{eqnarray}
\label{eq:rhov}
\rho^u_{n} & = & \frac{1}{2W^2u_p^n}\int d{\bf r} |\Delta{\bf u}({\bf r})|^n \nonumber\\
& = & \frac{1}{2W^2u_p^n}\int d{\bf r} |\mathcal{PT}{\bf u}({\bf r})-{\bf u}(\mathcal{P}{\bf r})|^n,
\end{eqnarray}
where $|\cdots|$ denotes the magnitude of a vector, $\mathcal{P}$ represents the appropriate odd or even parity operator, and $n>0$. Two other relevant variables that characterize the flow are the kinetic energy density field $\mathcal{E}({\bf r})=\varrho{\bf u}^2({\bf r})/2$, and the pseudoscalar vorticity field $\omega_z({\bf r})=\nabla\times{\bf u}({\bf r})$. The asymmetries in the kinetic energy density $\rho^\mathrm{KE}_O$ and $\rho^\mathrm{KE}_E$ are defined in a similar manner,
\begin{eqnarray}
\label{eq:rhoke}
\rho^{\mathrm{KE}}_{n} & = & \frac{1}{2W^2(\varrho u_p^2/2)^n}\int d{\bf r} |\Delta\mathcal{E}({\bf r})|^n\nonumber\\
& = & \frac{1}{2W^2 (\varrho u_p^2/2)^n}\int d{\bf r} |\mathcal{PT}\mathcal{E}({\bf r})-\mathcal{E}(\mathcal{P}{\bf r})|^n.
\end{eqnarray}
Due to the pseduoscalar nature of the vorticity field, its asymmetries $\rho^\omega_O$ and $\rho^\omega_E$ are defined with a positive sign,
\begin{eqnarray}
\label{eq:rhow}
\rho^\omega_{n} & = & \frac{1}{2W^2(u_p/w)^n}\int d{\bf r} |\Delta\omega_z({\bf r})|^n \nonumber\\
& = & \frac{1}{2W^2(u_p/w)^n}\int d{\bf r} |\mathcal{PT}\omega_z({\bf r})+\omega_z(\mathcal{P}{\bf r})|^n.
\end{eqnarray}
Notice that we have introduced $u_p$, $\varrho u_p^2/2$, and $u_p/w$ as the units of velocity, kinetic energy density, and vorticity respectively, and $W^2$ is the area of the flow domain. Eqs.(\ref{eq:rhov})-(\ref{eq:rhow}) are applicable for $\alpha\leq 1$. When $\alpha\geq 1$, the velocity unit changes to $\alpha u_p$ so that the equivalence between $\alpha\leftrightarrow 1/\alpha$ and the exchange of axes, $x\leftrightarrow y$, is preserved. Note that, by construction, the boundary contribution to the asymmetry in all $\mathcal{PT}$-symmetric configurations (Fig.~\ref{fig:PTSymmetryTypes}) is zero. In the next subsection, we describe the numerical method used to obtain the steady-state solution for the velocity field ${\bf u}({\bf r})$ in the presence of $\mathcal{PT}$-symmetric boundary conditions ${\bf u}_b({\bf r})$.


\subsection{Lattice Boltzmann method for viscous flow}
\label{subse:LBM}
In this work, we use two prevailing lattice Boltzmann models: the single-relaxation-time (SRT) model for 2D flow and the multiple-relaxation-time (MRT) model for 3D flow by using existing validated codes~\cite{re:Yu02,re:Yu05}. The corresponding lattice models are D2Q9~\cite{re:qian92} and D3Q19~\cite{rf:Qian97} respectively.

The SRT lattice Boltzmann equation for the D2Q9 lattice model~\cite{re:chen92,re:qian92} reads
\begin{equation}
f_{\beta}({\bf r}+{\bf e}_{\beta}\delta_{t},t+\delta_{t})= f_{\beta}({\bf r},t)-\frac{1}{\tau}\left[f_{\beta}({\bf r},t)-f_{\beta}^{\mathrm{eq}}({\bf r})\right]
\label{eq:LBE}
\end{equation}
where $f_{\beta} $ ($\beta=0,\cdots,8$) are the single-particle distribution functions, $f_\beta^{\mathrm{eq}}({\bf r})$ are the corresponding equilibrium distribution functions, $\delta_t$ is the time increment, $\delta_{\bf r}={\bf e}_\beta\delta_t$ is the incremental displacement of the lattice mesh, and $\tau$ is the dimensionless relaxation time, measured in units of $\delta_t$ and determined by molecular collisions. This relaxation time is related to the viscosity of the fluid. We use dimensions such that the ratio of spatial and temporal increments is unity, $c=\delta_x/\delta_t=1$. The discrete particle velocities ${\bf e}_\beta$ and the weighting factors $\omega_\beta$ are given by ${\bf e}_0=(0,0)c$ and $\omega_0=4/9$ for $\beta=0$, ${\bf e}_\beta=(\cos[(\beta-1)\pi/2],\sin[(\beta-1)\pi/2])c$ and $\omega_\beta=1/9$ for $\beta$=1-4, and ${\bf e}_\beta=(\cos[(\beta-4.5)\pi/2], \sin[(\beta-4.5)\pi/2])c$ and $\omega_\beta=1/36$ for $\beta$=5-8. The corresponding equilibrium distribution functions are given by~\cite{re:He97}
\begin{equation}
\label{eq:SRTLBE}
f_\beta^{\mathrm{eq}}({\bf r})=\omega_\beta
\left[\delta\varrho+\varrho_0\left(\frac{3{\bf e}_\beta\cdot{\bf u}}{c^2}+
\frac{9({\bf e}_\beta\cdot{\bf u})^2}{(2c^4)}-\frac{3{\bf u}^2}{(2c^2)}\right)\right].
\end{equation}
where $\delta\varrho$ is the density fluctuation, $\varrho_0$ is the constant mean density of the system, usually set to 1. The total density of the fluid is given by $\varrho=\delta\varrho+\varrho_0$.

The MRT lattice Boltzmann equation~\cite{re:dHumieres02} for the D3Q19 lattice model is given by
\begin{equation}
|f({\bf r}+{\bf e}_{\beta}\delta_{t},t+\delta_{t}\rangle= |f({\bf r},t)\rangle-M^{-1}\hat{S}\left(|m({\bf r},t)\rangle-|m^\mathrm{eq}({\bf r})\rangle\right),
\label{eq:MRTLBE}
\end{equation}
where the Dirac ket $|\cdots \rangle$ represents a column vector, thus $|f({\bf r},t)\rangle=\left[f_0({\bf r},t),\ldots,f_{18}({\bf r},t)\right]^T$. The discrete particle velocities ${\bf e}_\beta$ and the weighting factors $\omega_\beta$ ($\beta=0,\cdots,18$) are given by ${\bf e}_0=(0,0,0)c$ and $\omega_0=1/3$ for $\beta=0$, ${\bf e}_\beta=\left\{(\pm 1,0,0)c, (0,\pm 1,0)c, (0,0,\pm 1)c\right\}$ and $\omega_\beta=1/18$ for $\beta$=1-6, and ${\bf e}_\beta=\left\{(\pm 1,\pm 1,0)c, (\pm 1,0,\pm 1)c, (0,\pm 1,\pm 1)c\right\}$ and $\omega_\beta=1/36$ for $\beta$=7-18. The column vectors $|m({\bf r},t)\rangle$ and $|m^\mathrm{eq}({\bf r})\rangle$ represents the moments of distribution function $|f({\bf r},t)\rangle$ and the corresponding equilibrium distribution function $|f^\mathrm{eq}({\bf r})\rangle$, respectively. The diagonal dimensionless collision matrix $\hat{S}$ is given by $\hat{S}= \left(0,s_1,s_2,0,s_4,0,s_4,0,s_4,s_9,s_2,s_9,s_2,s_9,s_9,s_9,s_{16},s_{16},s_{16}
\right)$ where $s_2,s_4,s_9$ and $s_{16}$ are parameters corresponding to multiple relaxation time-scales. The details of the equilibrium moment vector $|m^\mathrm{eq}\rangle$, the transformation matrix $M$, and the diagonal matrix $\hat{S}$ for the D3Q19 lattice model can be found in Ref.~\cite{re:Yu06}.

The hydrodynamic variables are obtained via the moments of particle-distribution functions~\cite {re:Yu05,re:Yu06}
\begin{equation}
\delta\varrho({\bf r},t) =\sum_\beta f_\beta({\bf r},t), \quad \varrho_0{\bf u}({\bf r},t)=\sum_\beta{\bf e}_\beta f_\beta({\bf r},t),
\label{eq:hydro}
\end{equation}
The hydrodynamic pressure is given by $p=c_s^2\varrho$ where the speed of sound is $c_s=c/\sqrt{3}$ for both D2Q9 and D3Q19 lattice models. The kinematic viscosity is given by $\nu=(\tau-0.5)c^2\delta_t/3$ for the D2Q9 SRT model and $\nu=(s_9^{-1}-0.5)c^2\delta_t/3$ for the D3Q19 MRT model. It should be pointed out that the practice of using only $\delta\varrho$ instead of $\varrho$ in Eq.~(\ref{eq:hydro}) reduces the effects of round-off errors in the simulations~\cite{re:dHumieres02,re:Skordos93}. We specify identical parabolic (2D) and paraboloid (3D) velocity profiles with maximum velocity $u_p$ perpendicular to the cross-section and a constant pressure $p_0$ at the inlet and the outlet. The inlet velocity profile is introduced one grid before the inlet grid using the generalized bounce-back boundary condition which relates single-particle distribution functions $f_\beta$ for different discrete particle velocities~\cite{re:Luo00},
\begin{equation}
f_{\beta^*}=f_\beta-6\omega_\beta\varrho_0{\bf u}_b\cdot{\bf e}_\beta/c^2.
\label{eq:generalizedBounceBack}
\end{equation}
Here $\beta^*$ and $\beta$ are related by ${\bf e}_{\beta^*}=-{\bf e}_\beta$, and ${\bf u}_b({\bf r})$ denotes different velocity profiles characterized by $u_p$ or $\alpha u_p$ (Fig.~\ref{fig:schematic}). The walls of the the flow domain are regarded rigid and bounce-back boundary condition is imposed.

In the following two sections, we present the numerical results for the $\mathcal{PT}$ asymmetries and fluid-flow fields obtained via the lattice Boltzmann method. The inlet/outlet width is $\omega/W=0.1$ unless otherwise indicated.


\section{Flow Asymmetries in 2D $\mathcal{PT}$-symmetric Configurations}
\label{se:results}


\subsection{Power-law scaling}
\label{subse:power}

\begin{figure}[tbph]
\centering
\includegraphics[width=\columnwidth]{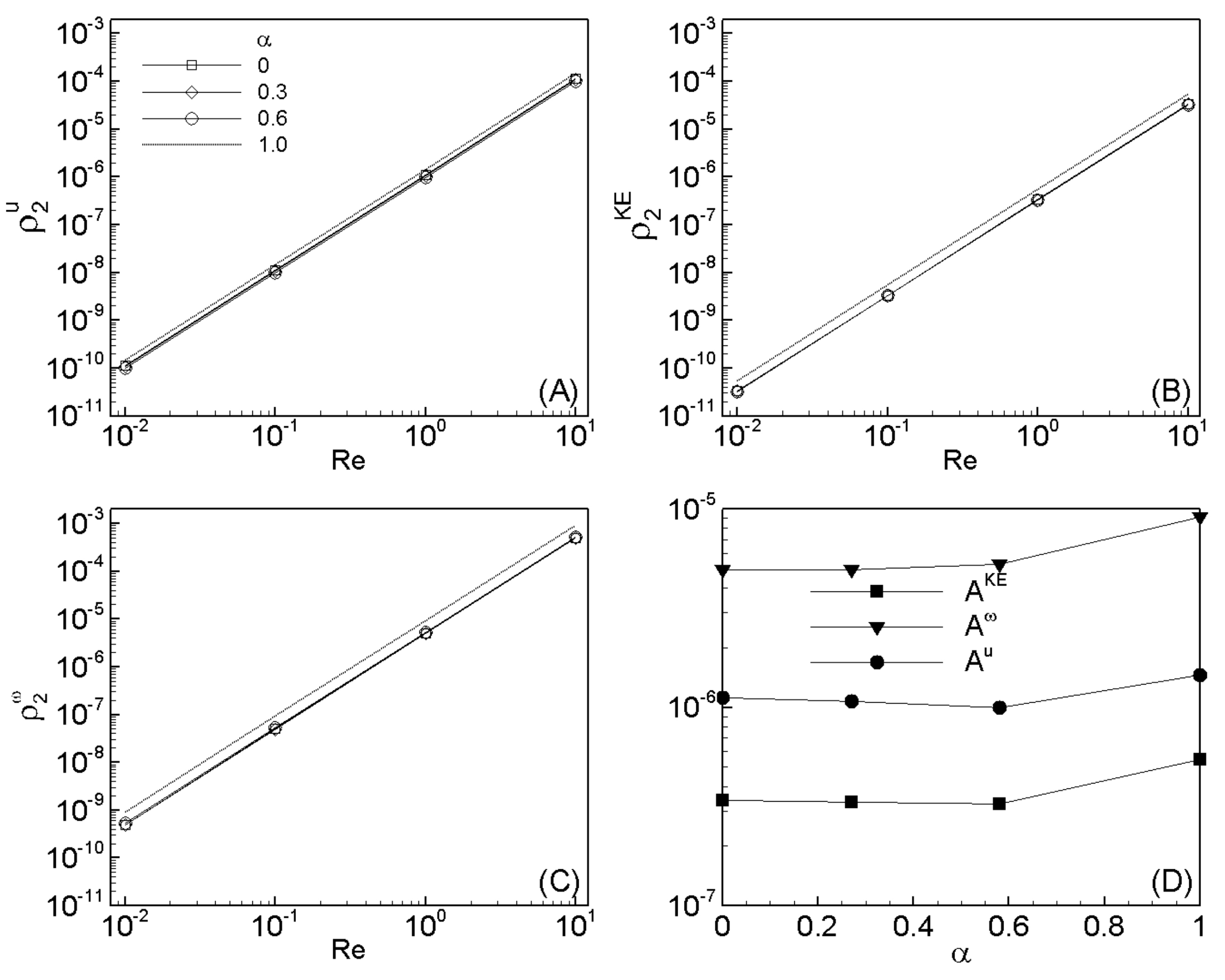}
\caption{Dependence of even-parity, $n=2$ asymmetries $\rho_2$ in velocity (A), kinetic energy density (B), and the vorticity (C) as a function of $\alpha$ shows that they scale quadratically with the Reynolds number, $\rho^i_2(\mathrm{Re},\alpha)=A^i(\alpha)\mathrm{Re}^2$ for $i=u,\mathrm{KE},\omega$.  Panel (D) shows the behavior of $A^i(\alpha)$ for $0\leq\alpha\leq 1$.}
\label{fig:rhoevenalpha}
\end{figure}

We first show the results for even-parity, $n=2$ asymmetries as a function of $(\mathrm{Re},\alpha)$ in Fig.~\ref{fig:rhoevenalpha} for the $\mathcal{PT}$-symmetric configuration in panel (b) of Figs.~\ref{fig:schematic} and~\ref{fig:PTSymmetryTypes}. It shows the dimensionless asymmetry in  velocity (A), kinetic energy density (B) and vorticity (C) as a function of Re over three decades for four different values of $0\leq\alpha\leq 1$; note the logarithmic scale on both axes. Panels (A), (B), and (C) shows that the $n=2$ asymmetries scale quadratically with $\mathrm{Re}$, $\rho^i_2(\mathrm{Re},\alpha)=A^i(\alpha)\mathrm{Re}^2$ where $i=u,\mathrm{KE},\omega$. Panel (D) shows that the prefactor $A^i(\alpha)$ increases with $\alpha$ for $0\leq\alpha\leq 1$. It should be noted that results for $\alpha\geq 1$ are obtained by exchanging the vertical and horizontal axes.

We emphasize here that these asymmetries, although small, are not numerical artifacts. The steady-state velocity field ${\bf u}({\bf r})$, numerically obtained via the LBM by using double-precision calculation, satisfies other symmetry constraints exceptionally well. For example, when $\alpha=0$ (vertical flow) reflection symmetry implies that the resultant velocity field must satisfy $u_x(x,y)=-u_x(-x,y)$ and $u_y(x,y)=u_y(-x,y)$. The dimensionless error in this constraint,
\begin{equation}
\label{eq:constraint}
\delta=\frac{1}{2W^2u_p}\int d{\bf r} \left\{\begin{array}{c} |u_x(x,y)+u_x(-x,y)| +\\
|u_y(x,y)-u_y(-x,y)|\end{array}\right\},
\end{equation}
satisfies $\delta \approx 10^{-30}$ for all Reynolds numbers considered in this paper. It is also noted that that the integrands for the asymmetries $\rho^u_2$ and $\rho^\mathrm{KE}_2$ contain two distinct powers - second and fourth, respectively - of the steady-state velocity field. {\it This quadratic scaling of the even-parity, $n=2$ asymmetries is the first significant result of this paper.}

\begin{figure}[tbhp]
\centering
\includegraphics[width=\columnwidth]{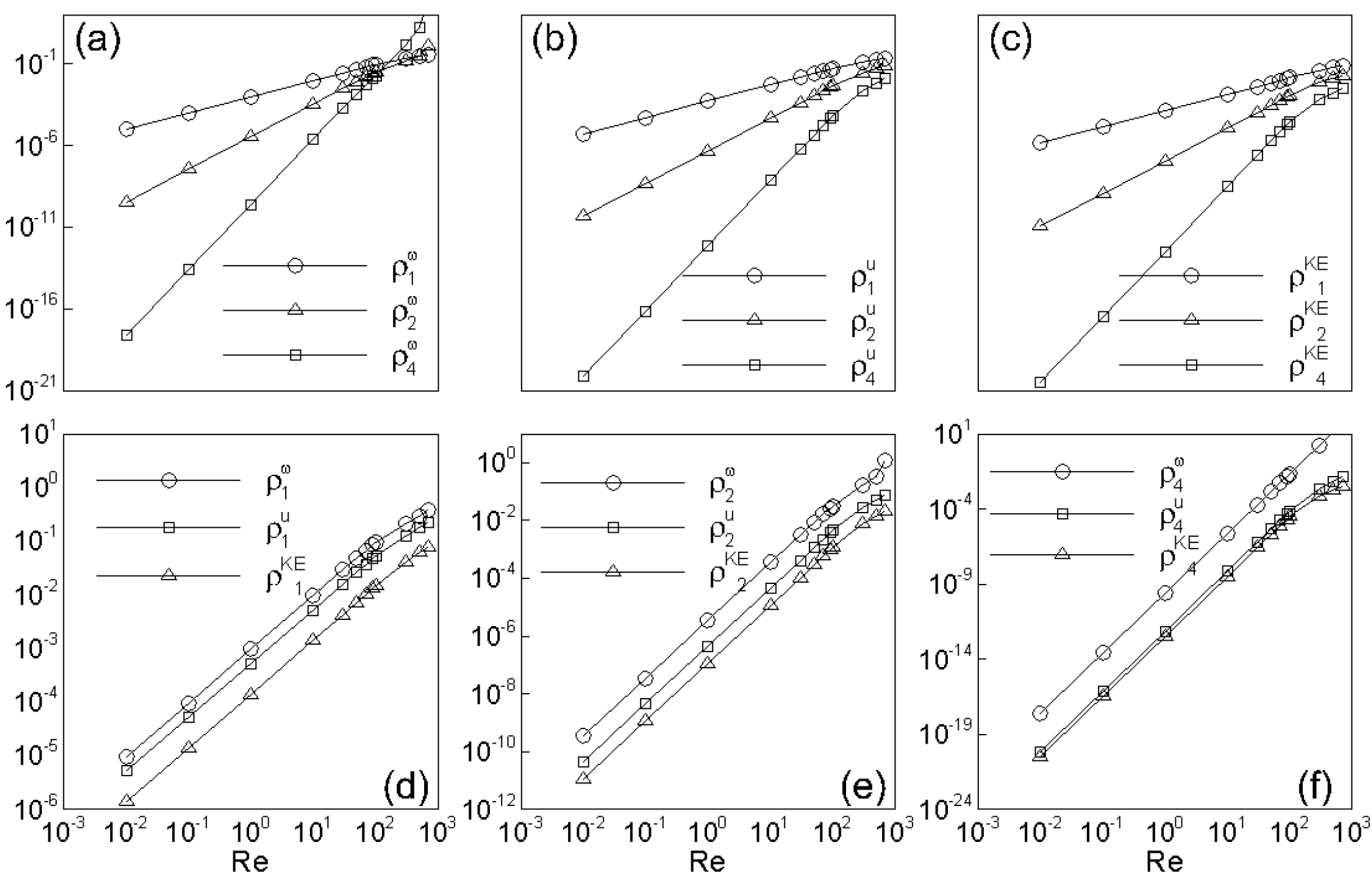}
\caption{For panel (b) of Fig.~\ref{fig:PTSymmetryTypes} which satisfies both even- and odd-parity asymmetries, the dependence of $\mathcal{PT}$-asymmetry in vorticity (a), velocity (b),and kinetic energy density (c) as a function of Re for $n=\{1, 2, 4\}$ respectively shows that they scale the $n-{th}$ power law with the Reynolds number, $\rho^i(\mathrm{Re},n)=A^i(n)\mathrm{Re}^n$ for $i=u,k,\omega$. Each index of $n=1$ in (d), $n=2$ in (e), and $n=4$ in (f) for velocity ($u$), kinetic energy density($KE$), and vorticity ($\omega$) shows the power law scaling is independent to the fields.}
\label{fig:threerho}
\end{figure}
To investigate the origin of the quadratic power-law scaling, we compare the linear ($n=1$), quadratic ($n=2$), and quartic ($n=4$) asymmetries $\rho^i_n(\mathrm{Re})$ ($i=u,\mathrm{KE},\omega$) for $\alpha=1$ in the range of $\mathrm{Re}=0.01-700$. The reader is reminded that when $\alpha=1$, due to the exact reflection symmetry across the northeast-to-southwest diagonal of the square, the velocity profile satisfies $u_x(x,y)=u_y(y,x)$ and $u_y(x,y)=u_x(y,x)$, and therefore, the odd- and even-parity asymmetries are identical in this configuration. The top row in Fig.~\ref{fig:threerho} shows the three asymmetries for vorticity (panel a), velocity (panel b), and kinetic energy density (panel c). The vertical scale in each panel in the top row is the same, and the horizontal scale is identical to that in the bottom row. It is clear that all asymmetries scale algebraically with the Reynolds number with an exponent equal to $n$, $\rho^i_n(\mathrm{Re},\alpha=1)=B^i_n\mathrm{Re}^n$ for $i=u,\mathrm{KE},\omega$.  The bottom row in Fig.~\ref{fig:threerho} displays the velocity, kinetic energy density, and vorticity asymmetries for $n=1$ (panel d), $n=2$ (panel e), and $n=4$ (panel f). It shows clearly that the $\mathcal{PT}$ asymmetries $\rho^i$ ($i=u,\mathrm{KE},\omega$) exhibit identical power-law scaling with the Reynolds number over four decades,
\begin{equation}
\label{eq:scaling}
\rho^i_n(\mathrm{Re},\alpha)=A^i_n(\alpha) \mathrm{Re}^n.
\end{equation}
{\it This $n$-dependent power-law scaling is our second significant result.}

\begin{figure}[tbph]
\centering
\includegraphics[width=0.6\columnwidth]{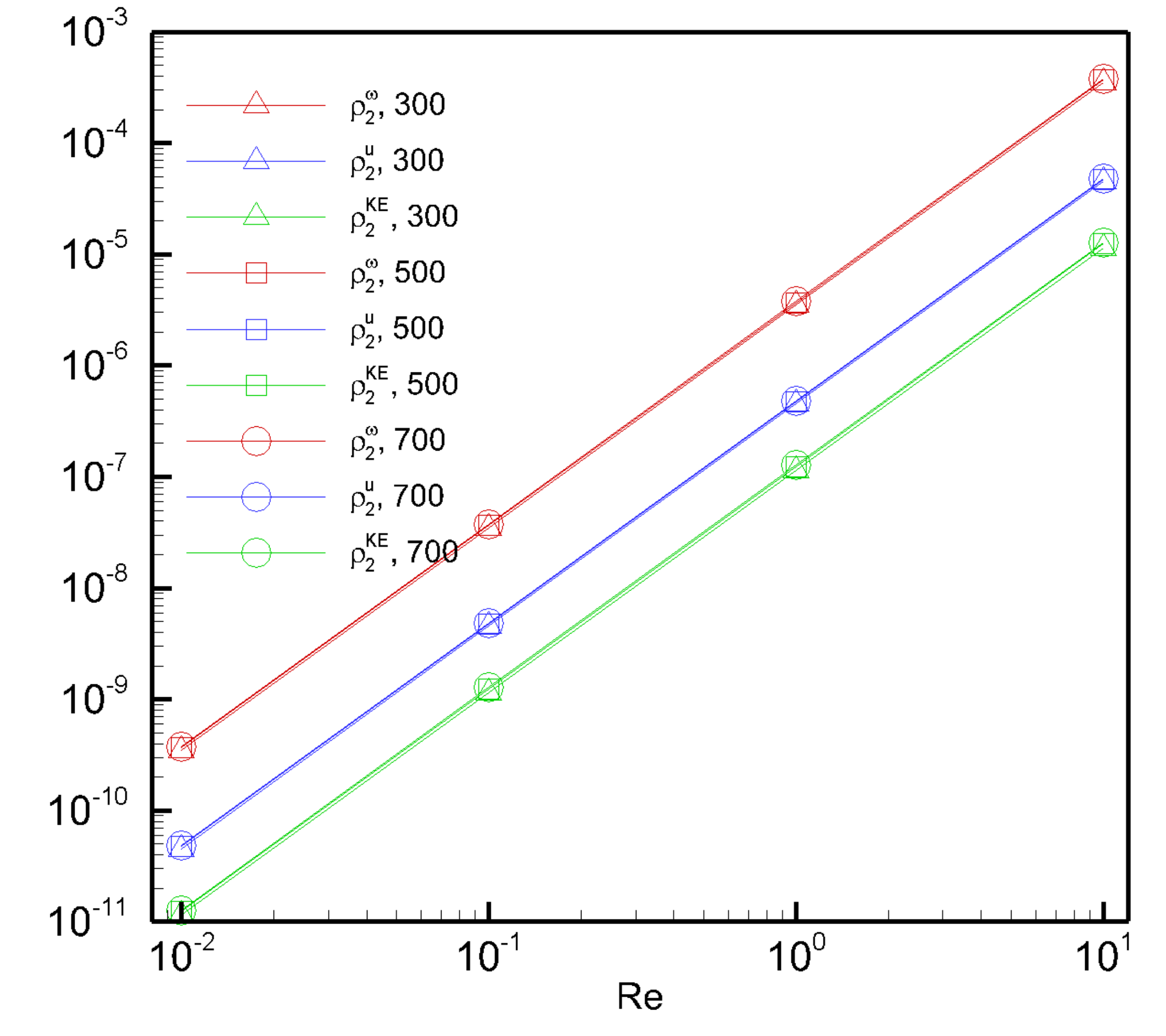}
\caption{Comparison of scaling behavior of asymmetries $\rho_2^i=A_2^i(\alpha=1)\mathrm{Re}^2$ in velocity (blue), vorticity (red), and kinetic energy density (green) for three grid-resolutions. The virtually identical results for grid-size $300^2$ ($\Delta$), $500^2$ ($\Box$), and $700^2$ ($\bigcirc$) shows that the resolution of $300^2$, used for 2D square geometries in this paper, are independent of discretization.}
\label{fig:ResolutionCheck}
\end{figure}
To verify that these results are independent of the grid-discretization used in the LBM, we perform space resolution convergence check for the 2D square case, Fig.~\ref{fig:PTSymmetryTypes}(a) with $\alpha=1$, to determine the optimal discretization. We use three grid resolutions, $300^2$ ($\Delta$), $500^2$ ($\Box$), and $700^2$ ($\bigcirc$) to generate the steady-state velocity fields for Reynold numbers ranging over three orders of magnitude, $\mathrm{Re}=\{0.01, 0.1, 1, 10\}$. Figure~\ref{fig:ResolutionCheck} shows the second-order asymmetries in velocity $\rho_2^u$ (blue),
vorticity $\rho_2^\omega$ (red), and kinetic energy $\rho_2^{\mathrm{KE}}$ (green) as a function of Reynolds number on logarithmic scale. It is clear, from the complete overlap of the results obtained via different resolutions that a resolution of $300^2$ is sufficient for the 2D results presented here.

The scaling behavior, encapsulated in Figs.~\ref{fig:rhoevenalpha} and~\ref{fig:threerho}, raises two questions. Why are the asymmetries $\rho^i$ characterized by a {\it single exponent} over four decades in Reynolds numbers that span from $\mathrm{Re}\ll 1$ to $\mathrm{Re}\gg 1$? Why do the  asymmetries in velocity field and kinetic energy density - which depends quadratically on the velocity field - have the same power-law exponent? To answer these questions, note that in the limit $\mathrm{Re}=0$, corresponding to a Stokes flow, a balanced inflow-outflow configuration results in a $\mathcal{PT}$-symmetric flow velocity field ${\bf u}_{S}({\bf r})$. When $\mathrm{Re}\neq 0$, the solution of the NS equation ${\bf u}({\bf r})$ can be expressed as a sum of the Stokes flow ${\bf u}_S({\bf r})$ and a correction term ${\bf u}_{A}({\bf r})$. The Taylor-series expansion of the correction term starts at the first order in Reynolds number,
\begin{equation}
\label{eq:sas}
{\bf u}_A({\bf r})=\mathrm{Re} {\bf u}_1({\bf r})+\mathrm{Re}^2 {\bf u}_2({\bf r})+\mathrm{Re}^3 {\bf u}_3({\bf r})+\ldots,
\end{equation}
where the vector fields ${\bf u}_k({\bf r})$ are not necessarily $\mathcal{PT}$-symmetric. It follows that the asymmetries in the velocity $\Delta{\bf u}_A(\bf r)=\mathcal{PT}{\bf u}_A({\bf r})-{\bf u}_A(\mathcal{P}{\bf r})$ and kinetic energy density $\Delta\mathcal{E}({\bf r})$ can be written as
\begin{eqnarray}
\label{eq:duv}
\Delta{\bf u}_A({\bf r})& = & \mathrm{Re}\left[\mathcal{PT}{\bf u}_1({\bf r})-{\bf u}_1(\mathcal{P}{\bf r})\right]+\mathrm{Re}^2\left[\mathcal{PT}{\bf u}_2({\bf r})-{\bf u}_2(\mathcal{P}{\bf r})\right]+\ldots,\nonumber\\
2\frac{\Delta{\mathcal E}({\bf r})}{\varrho} & = & 2{\bf u}_S({\bf r})\cdot\Delta{\bf u}_A({\bf r})+\left[\mathcal{PT}{\bf u}^2_A({\bf r})-{\bf u}_A^2(\mathcal{P}{\bf r})\right],
\end{eqnarray}
with a corresponding expression for vorticity as well. It follows from Eqs.(\ref{eq:duv}) and (\ref{eq:rhov}) that at small values of $\mathrm{Re}\ll 1$, the asymmetries in velocity and vorticity must scale with a single power-law exponent, $\rho_n^u\propto\mathrm{Re}^n$. It also follows from Eq.(\ref{eq:duv}) that in the same regime $\Delta\mathcal{E}({\bf r})$ is also linear in the Reynolds number and therefore, the kinetic energy asymmetry scales with the same exponent, $\rho_n^\mathrm{KE}\propto\mathrm{Re}^n$.

We emphasize that these considerations are valid only for small Reynolds numbers and at $\mathrm{Re}\gtrsim 1$, higher-order terms in the Taylor series expansion, Eq.(\ref{eq:sas}), are expected to become relevant. The contribution from these terms is expected to change the asymmetry scaling from a single-exponent scaling to a polynomial scaling in the Reynolds number. However, Fig.~\ref{fig:threerho} shows that {\it the single-exponent scaling remains valid at significantly higher values of Reynolds numbers, $\mathrm{Re}\sim 100$.} The origin of this robustness remains the subject of ongoing investigation.

\subsection{Balanced vs. fully developed outflow}
\label{subse:fdo}

\begin{figure}[htpb]
\centering
\includegraphics[width=\columnwidth]{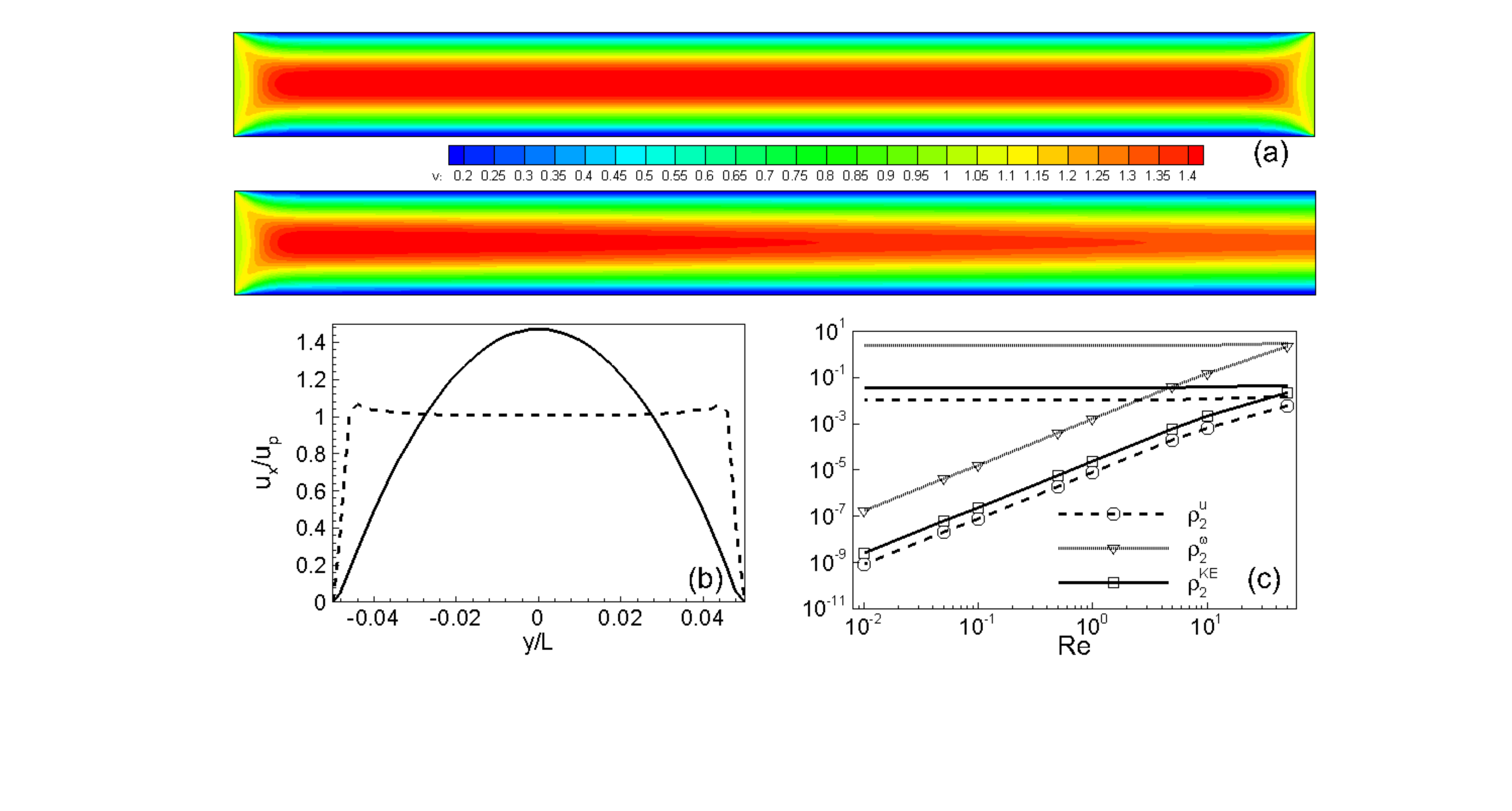}
\vspace{-15mm}
\caption{Comparison of fluid flows with $\mathcal{PT}$-symmetric (top color map) or fully-developed-outflow (bottom color map) constraints. The geometry and velocity profile at the inflow is identical in both cases. Panel (a) shows the steady-state fluid speed $|{\bf u}({\bf r})|$ in a long channel with $L/W=10$, a wide inlet $w/W=1$, and Reynolds number $\mathrm{Re}=0.1$. We see that the flow structure near the outlet is dramatically different for the two configurations. Panel (b) shows that the horizontal fluid-velocity component $u_x(L/2,y)$ for the fully developed outflow (dashed line) is almost constant with a local minimum at the center, $y=0$. Panel (c) shows that the $n=2$ asymmetries scale as $\mathrm{Re}^2$ for the $\mathcal{PT}$-symmetric configuration (lines with open symbols), whereas asymmetries for the fully-developed-outflow configuration (lines with solid symbols) are virtually constant as the Reynolds number changes over three-and-a-half decades.}
\label{fig:fdo}
\end{figure}
In this subsection, we explore the universality of the power-law scaling expressed in Eq.~(\ref{eq:scaling}) to see if it depends upon the boundary velocity profile ${\bf u}_b({\bf r})$, or the square geometry, or the $\mathcal{PT}$-symmetric inflow-outflow instrumental for it. Calculations carried out with three different inflow/outflow velocity profiles, uniform, parabolic, and triangular, for a horizontal flow in a square domain, i.e. panel (b) ($1/\alpha=0$) in Figs. \ref{fig:PTSymmetryTypes}, show that the $n=2$ asymmetries in all variables scale quadratically with the Reynolds number, although the prefactor $A^i_2$ in Eq.~(\ref{eq:scaling}) varies slightly ~\cite{china}.

To investigate the dependence of power-law scaling on the flow-domain geometry and $\mathcal{PT}$-symmetric boundary conditions, we consider a long horizontal channel with length $L=10W$, open inlet/outlet with $w/W=1$, and uniform stream with velocity $u_p$ from left to right at inlet. The Reynolds number $\mathrm{Re}=1$. At the outlet, we impose two different boundary conditions: the first one is identical to the inflow, satisfying the $\mathcal{PT}$-symmetry and the second one is fully-developed boundary condition, meaning there is no velocity gradient at the outlet, $\partial_x{\bf u}(x=L/2,y)=0$. It is pointed that in laminar flow regime when the flow is fully developed the second flow develops a parabolic velocity profile downstream ~\cite{landau}.

We show results on the steady-state velocity field in Fig.~\ref{fig:fdo} from three aspects to compare the two outlet boundary conditions. Panel (a) shows the downstream velocity contours of $\mathcal{PT}$-symmetric outflow constraint (top color map) and the fully-developed-outflow constraint (bottom color map). It is clear that the $\mathcal{PT}$ asymmetries will be larger for the fully-developed-outflow constraint because of the uncorrelated inflow and outflow. Panel (b) shows a downstream velocity profile normalized by the uniform inflow near the outlet, $u_x(496L/500,y)/u_p$, as a function of $y$-coordinate in units of $L$. Note that we display results {\it near the boundary} instead of {\it at the boundary} since the velocity profiles at the boundary are fixed by the constraints. As is expected, the velocity profile is nearly uniform for the $\mathcal{PT}$-symmetric configuration (dashed line), whereas it is parabolic with a centerline velocity $3u_p/2$ for the fully-developed-outflow configuration (solid line). The dependence of $n=2$ $\mathcal{PT}$ asymmetries $\rho^i_2$ ($i=u,\mathrm{KE},\omega$), Eqs.(\ref{eq:rhov})-(\ref{eq:rhow}), on the Reynolds number over three-and-a-half decades is shown in panel (c). The dotted, dashed, and solid lines denote asymmetries in vorticity, kinetic energy density, and velocity respectively. The lines with open symbols show that, even for a rectangular geometry and a uniform inflow velocity profile, the $n=2$ asymmetries scale quadratically for a balanced inflow-outflow configuration. The asymmetries for the fully developed outflow, on the other hand, are shown by {\it the virtually flat lines with solid symbols.} Thus, the asymmetries in the fully-developed-outflow configuration are essentially independent of the Reynolds number, and their approximately constant value scales inversely with the aspect ratio $L/W$. Thus, our results show that $\mathcal{PT}$-symmetric configuration is instrumental to the power-law scaling of asymmetries $\rho^i_n$ ($i=u,\mathrm{KE},\omega$). In particular, the asymmetries in the balanced configuration at low Reynolds numbers are orders of magnitude smaller than those in the traditional, fully-developed-outflow configuration.

These results show that $\mathcal{PT}$-symmetric inflow-outflow configurations strongly suppress {flow fields'} asymmetries compared to their traditional counterparts. Since the dimensionless asymmetries $\rho^i_n$ represent integrated contributions, they do not possess information about their local structure. In the next subsection, we present the steady-state velocity field ${\bf u}({\bf r})$, kinetic energy density $\mathcal{E}({\bf r})$ and vorticity $\omega_z({\bf r})$ as a function of the Reynolds number. As we will show below, this detailed view provides further insights into the dramatic difference between $\mathcal{PT}$-asymmetries in the balanced inflow-outflow configuration and the more traditional, fully-developed-outflow configuration.


\subsection{Emergence of asymmetry in 2D flow patterns}
\label{subse:field}
\begin{figure}[htpb]
\centering
\includegraphics[width=0.47\textwidth]{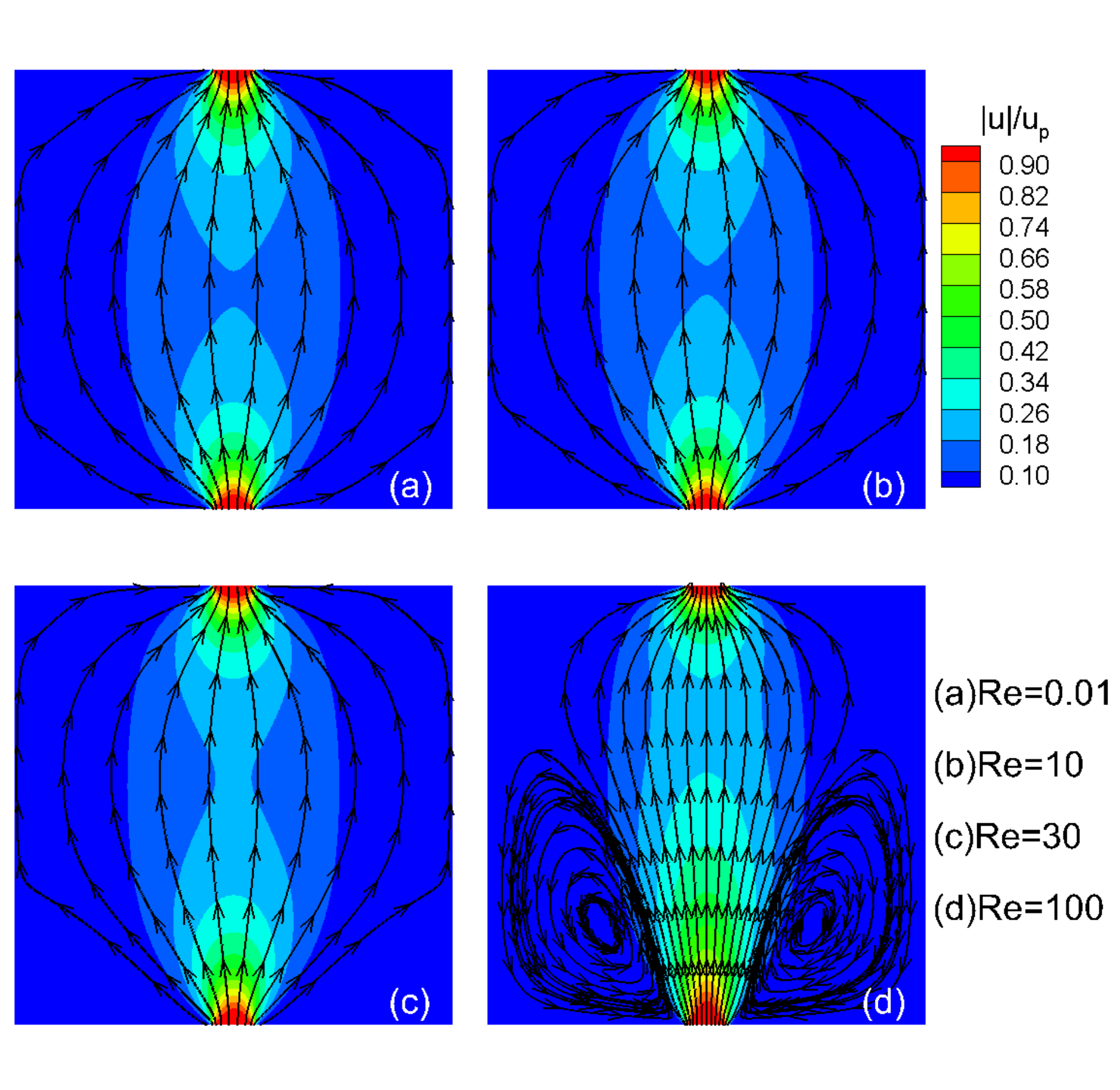}
\includegraphics[width=0.50\textwidth]{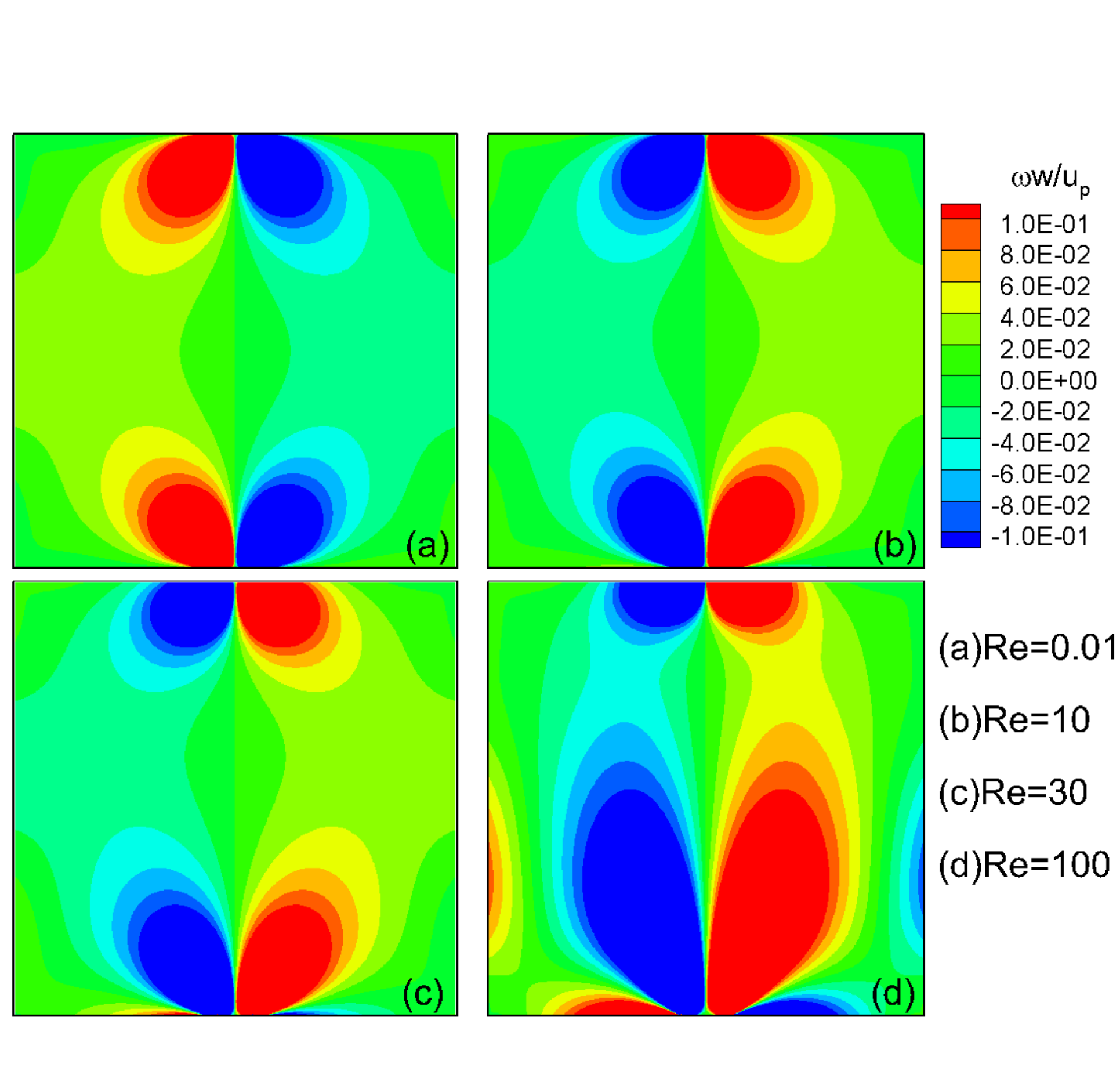}
\caption{Reynolds-number dependence of the velocity field ${\bf u}({\bf r})$ (left-hand side) and the vorticity field $\omega_z({\bf r})$ (right-hand side) for an $\alpha=0$, balanced inflow-outflow configuration. When $\mathrm{Re}=0.01$, panel (a), the velocity and vorticity profiles are essentially symmetrical about the diagonal; for $\mathrm{Re}=10$, panel (b), and $\mathrm{Re}=30$, panel (c), the asymmetry is visible but not prominent. When $\mathrm{Re}=100$, panel (d), the asymmetry is accentuated by the emergence of vortices near the inflow region that are absent near the outflow region. Note that the net vorticity in panels (a)-(d) on the right-hand side vanishes, $\Omega_z=0$.}
\label{fig:velvortalpha0}
\end{figure}

We start with the Re-dependence of velocity and vorticity fields in a square geometry with $\alpha=0$ corresponding to panel (c) in Fig.~\ref{fig:PTSymmetryTypes}. The left-hand side of Fig.~\ref{fig:velvortalpha0} shows the dependence of the steady-state velocity field ${\bf u}({\bf r})$, in the presence of $\mathcal{PT}$-symmetric inflow-outflow conditions, as a function of Reynolds number. This configuration is $\mathcal{PT}$-symmetric where parity corresponds to reflection in the horizontal axis. When the Reynolds number is quite small, $\mathrm{Re}=0.01$ (panel a), the velocity field is approximately $\mathcal{PT}$ symmetric. As the Reynolds number increases to $\mathrm{Re}=10$ (panel b) and $Re=30$ (panel c), however, the asymmetry in the velocity field at point ${\bf r}$ and its parity counterpart ${\bf r}_E=-{\bf r}$ are clearly present. (Recall that the origin of the coordinate system is at the center of the square.) In particular, when $\mathrm{Re}=100$, panel (d), vortices form near the inlet, but are absent near the outlet. Thus, panels (a)-(d) on the left-hand side of Fig.~\ref{fig:velvortalpha0} elucidate the origin of velocity asymmetries $\rho^u_n(\mathrm{Re})$ in a balanced inflow-outflow configuration. The right-hand side of Fig.~\ref{fig:velvortalpha0} shows the corresponding evolution of the pseudoscalar vorticity field $\omega_z({\bf r})$. It is seen that starting from an approximately $\mathcal{PT}$-symmetric result at small Reynolds number $\mathrm{Re}=0.01$, panel (a), the vorticity field, too, develops a strong asymmetry at relatively large Reynolds number $\mathrm{Re}=100$, panel (d). This asymmetry is primarily due to the presence of vortices near the inlet and their concomitant absence near the outlet.
\begin{figure}[htpb]
\centering
\includegraphics[width=0.50\textwidth]{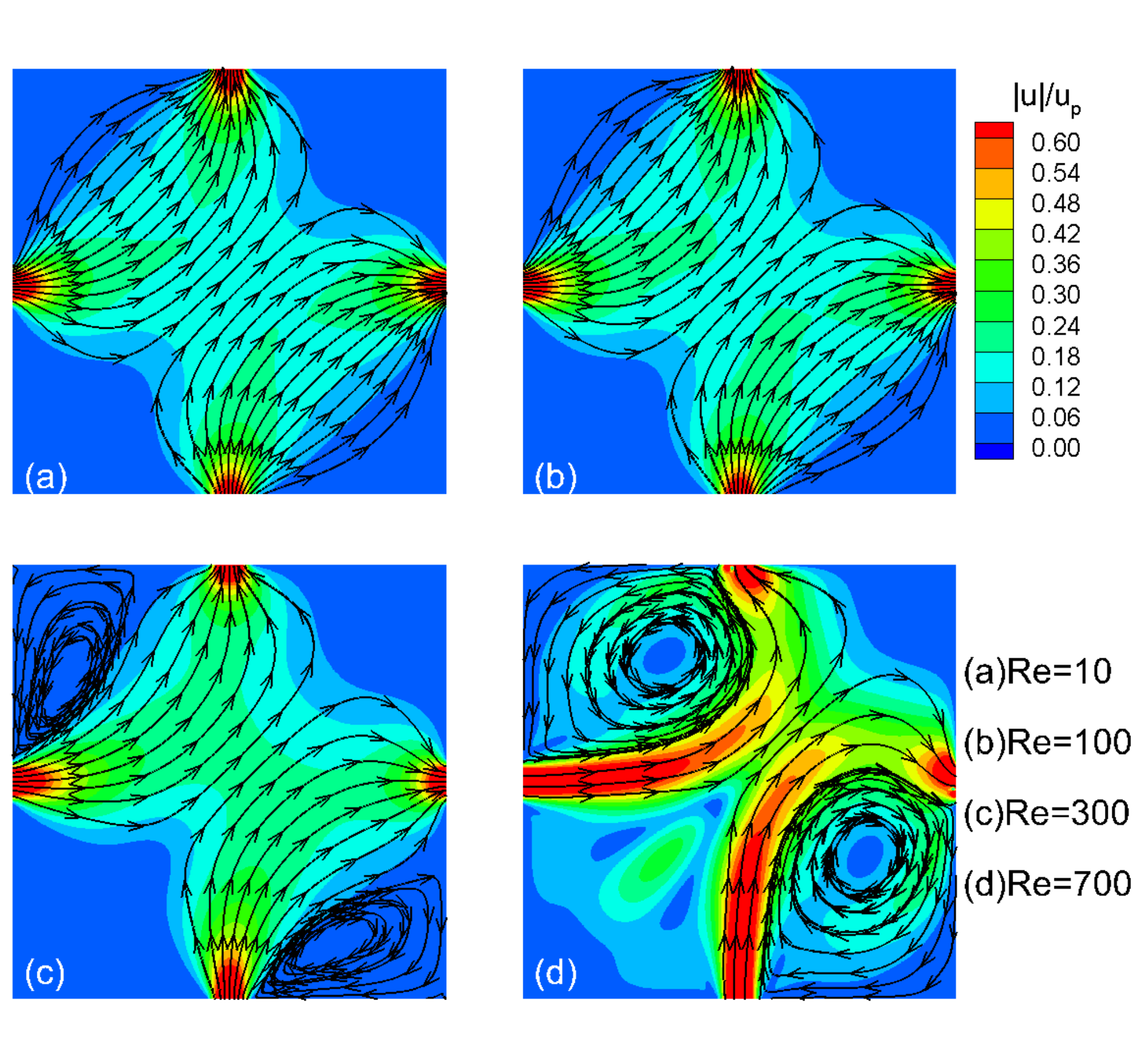}
\includegraphics[width=0.47\textwidth]{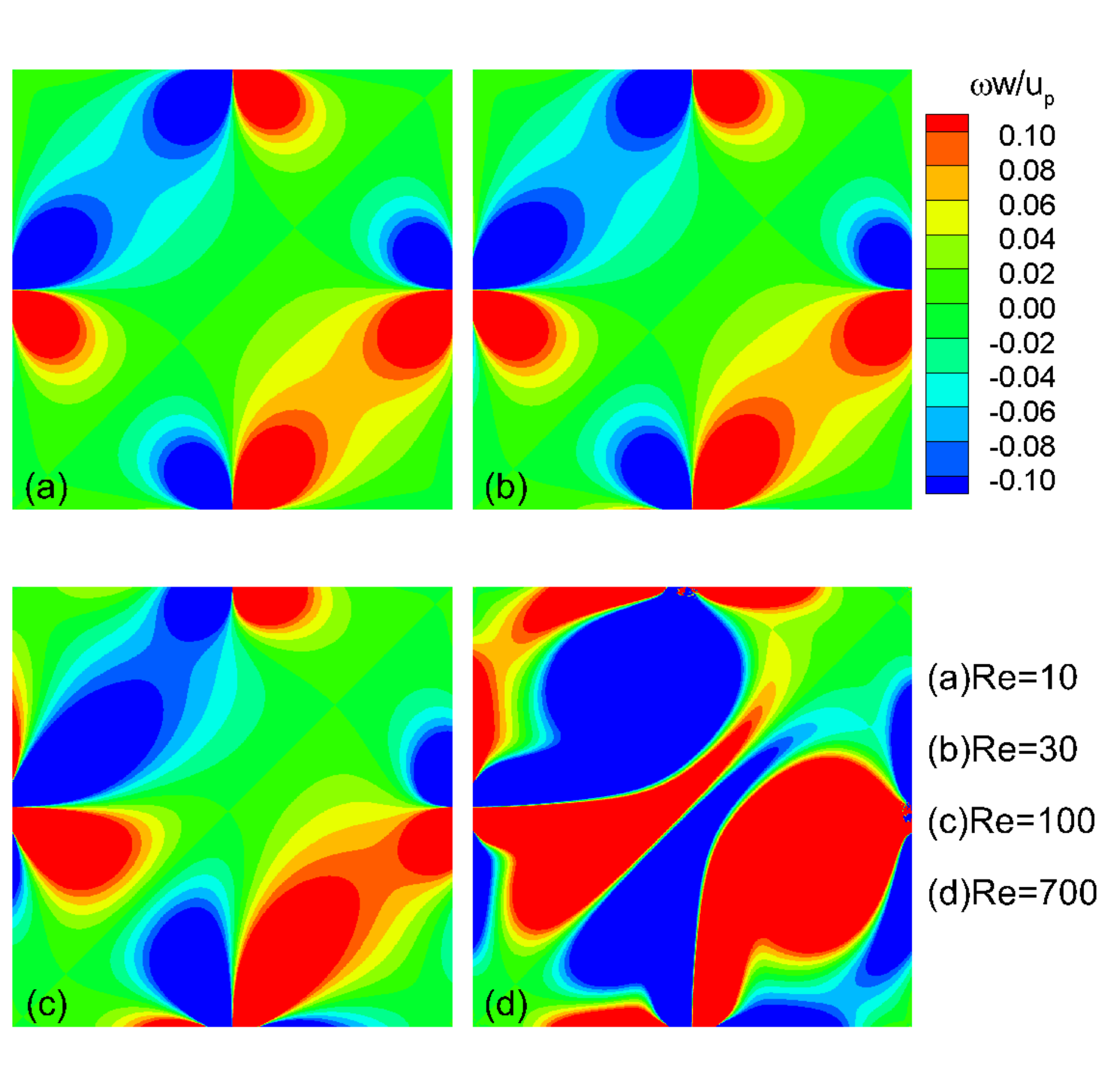}
\caption{Reynolds-number dependence of the velocity field ${\bf u}({\bf r})$ (left-hand side) and the vorticity field $\omega_z({\bf r})$ (right-hand side) for an $\alpha=1$, balanced inflow-outflow configuration. When $\mathrm{Re}=10$, panel (a), the velocity and vorticity profiles are essentially symmetrical about the horizontal axis; for $\mathrm{Re}=100$, panel (b) the asymmetry is visible but not prominent. When $\mathrm{Re}=300$, panel (c), and $\mathrm{Re}=700$, panel (d), the asymmetry is accentuated by the emergence of vortices near the inflow region that are absent near the outflow region. We point out that the color-scale in the velocity map is chosen to emphasize the low-velocity features; the maximum value of $|{\bf u}|/u_p$ is one.}
\label{fig:velvortalpha1}
\end{figure}

We note that the $\alpha=0$ velocity field is symmetric about the vertical axis, $u_x(x,y)=-u_x(-x,y)$ and $u_y(x,y)=u_y(-x,y)$, Eq.(\ref{eq:constraint}). Therefore, the vorticity field satisfies $\omega_z(x,y)=-\omega_z(-x,y)$ and the net vorticity is zero, $\Omega_z=W^{-2}\int d{\bf r}\omega_z({\bf r})=0$. Our numerical results satisfy this constraint exceptionally well. We find that $|\Omega_z|w/u_p< 10^{-17}$ for all Reynolds numbers that are considered.

Figure~\ref{fig:velvortalpha1} shows similar results for the velocity and vorticity fields for a balanced inflow-outflow configuration with $\alpha=1$, panel (d) in Fig.~\ref{fig:PTSymmetryTypes}. The left-hand side shows that the velocity field ${\bf u}({\bf r})$ develops asymmetries, accompanied by the emergence of vortices near the inflow region, as the Reynold number increases. Similarly, the right-hand side shows that the vorticity field asymmetries, too, grow with the Reynolds number. When $\alpha=1$, the velocity field is symmetrical about the southwest-to-northeast diagonal and satisfies $u_x(x,y)=u_y(y,x)$. Therefore, the vorticity field is antisymmerical about the same diagonal, and satisfies $\omega_z(x,y)=-\omega_z(y,x)$. Our numerical results satisfy the former to an accuracy of $\delta<10^{-30}$ and the latter to an accuracy of $|\Omega_z|w/u_p<10^{-17}$.

\begin{figure}[hbtp]
\centering
\includegraphics[width=0.48\textwidth]{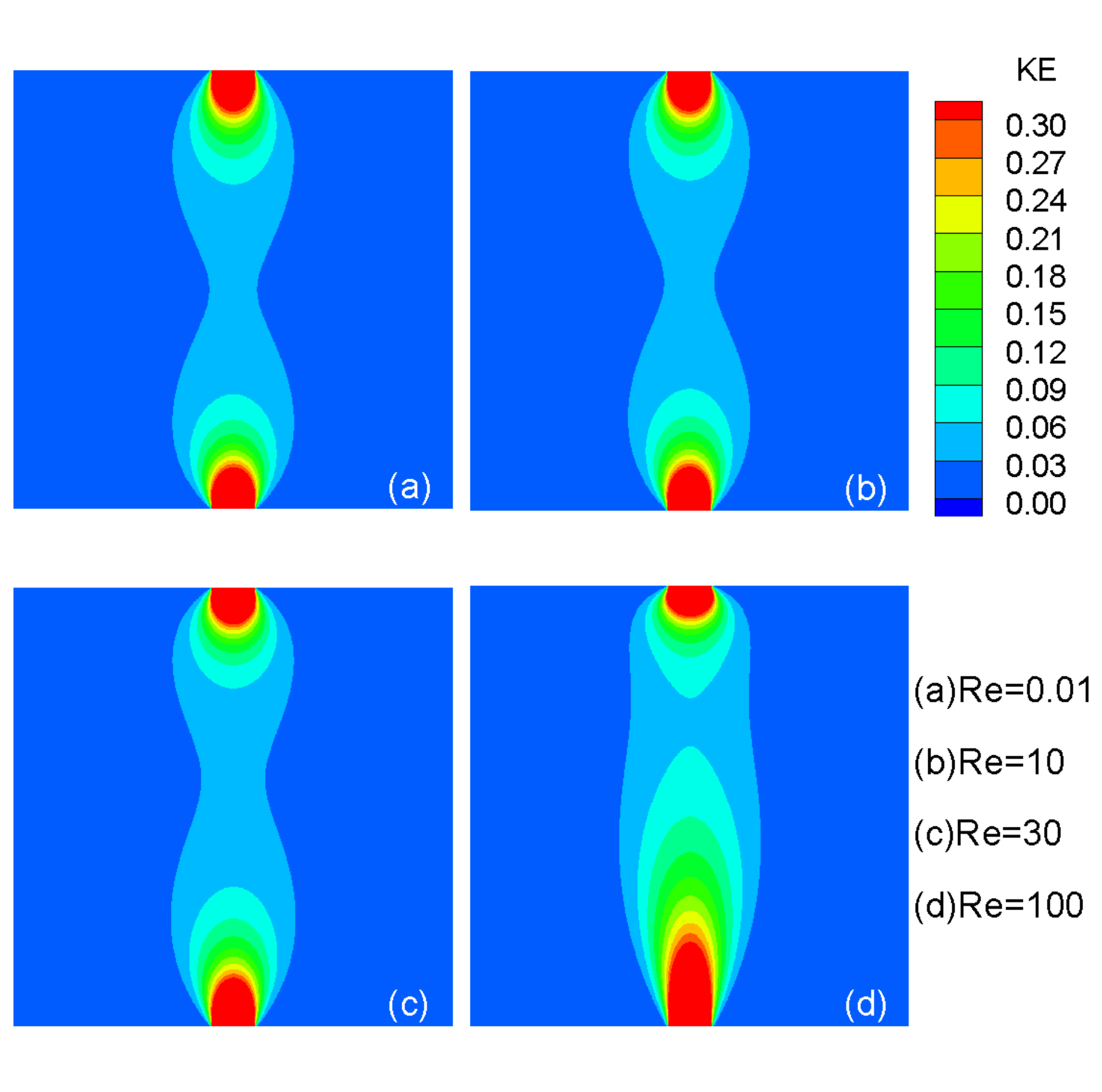}
\includegraphics[width=0.51\textwidth]{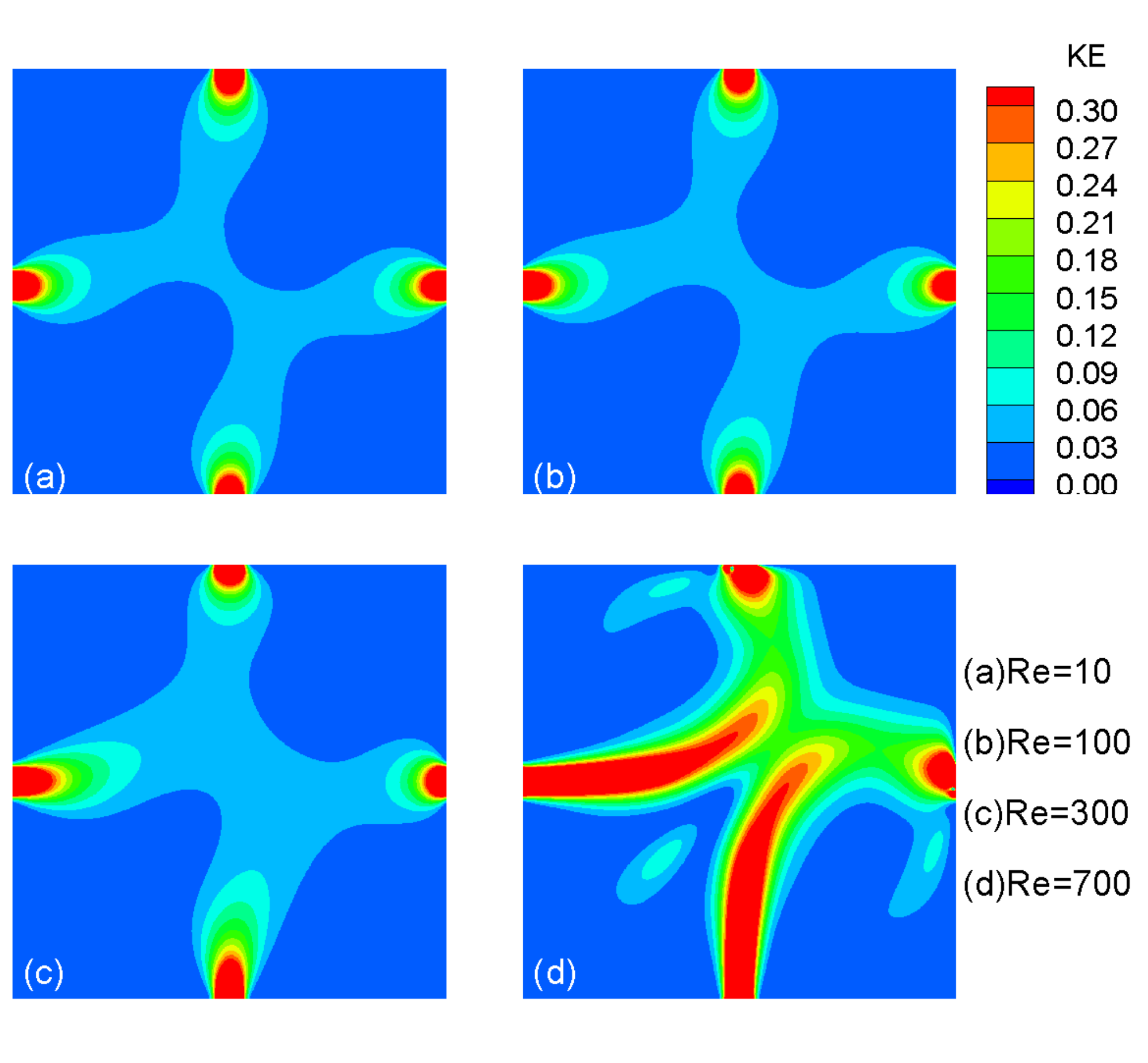}
\caption{Dependence of kinetic energy density on Reynolds number vertical flow (left-hand side)
and $\alpha=1$ flow (right-hand side). The energy density is scaled by $\varrho u_p^2/2$ and the color-map scale is chosen to emphasize the low-energy features; the maximum value of scaled energy density is one. Both cases show that the small $\mathcal{PT}$-asymmetry at low $\mathrm{Re}$, panel (a), is enhanced at large $\mathrm{Re}$, panel (d), due to the presence of vortices near the inlet and their absence near the outlet.}
\label{fig:kevsre}
\end{figure}
Figure~\ref{fig:kevsre} shows the kinetic energy density distributions as a function of Reynold number. The left-hand side shows that, for a vertical flow, as the Reynolds number increases from $\mathrm{Re}=0.01$, panel (a), to $\mathrm{Re}=100$, panel (d), the asymmetry $\rho^\mathrm{KE}_n$ about the horizontal axis increases. The right-hand side shows that for the $\alpha=1$ case, the asymmetry about the northwest-to-southeast diagonal increases. When $\mathrm{Re}=10$, panel (a), the kinetic energy distribution is almost $\mathcal{PT}$-symmetric; as the Reynold number increases to $\mathrm{Re}=700$, panel (d), the asymmetry is clearly visible. These results show that the kinetic energy asymmetries in $\mathcal{PT}$-symmetric configurations are driven by vortex formation near the inlet and its absence near the outlet.


\section{Power-law scaling of $\mathcal{PT}$ asymmetries in 3D: preliminary results}
\label{se:3d}
The power-law scaling of asymmetries in 2D $\mathcal{PT}$-symmetric configurations, expressed in Eq. (~\ref{eq:scaling}), is observed in the laminar regime crossing four decades of the Reynolds number. Numerical results presented in Sec.~\ref{se:results} strongly suggest that power-law scalings are universal in two-dimensional laminar flows of balanced inflow-outflow configurations. In this section, we present preliminary results for the same in three dimensions.

\begin{figure}[htbp]
\centering
\includegraphics[width=0.7\columnwidth]{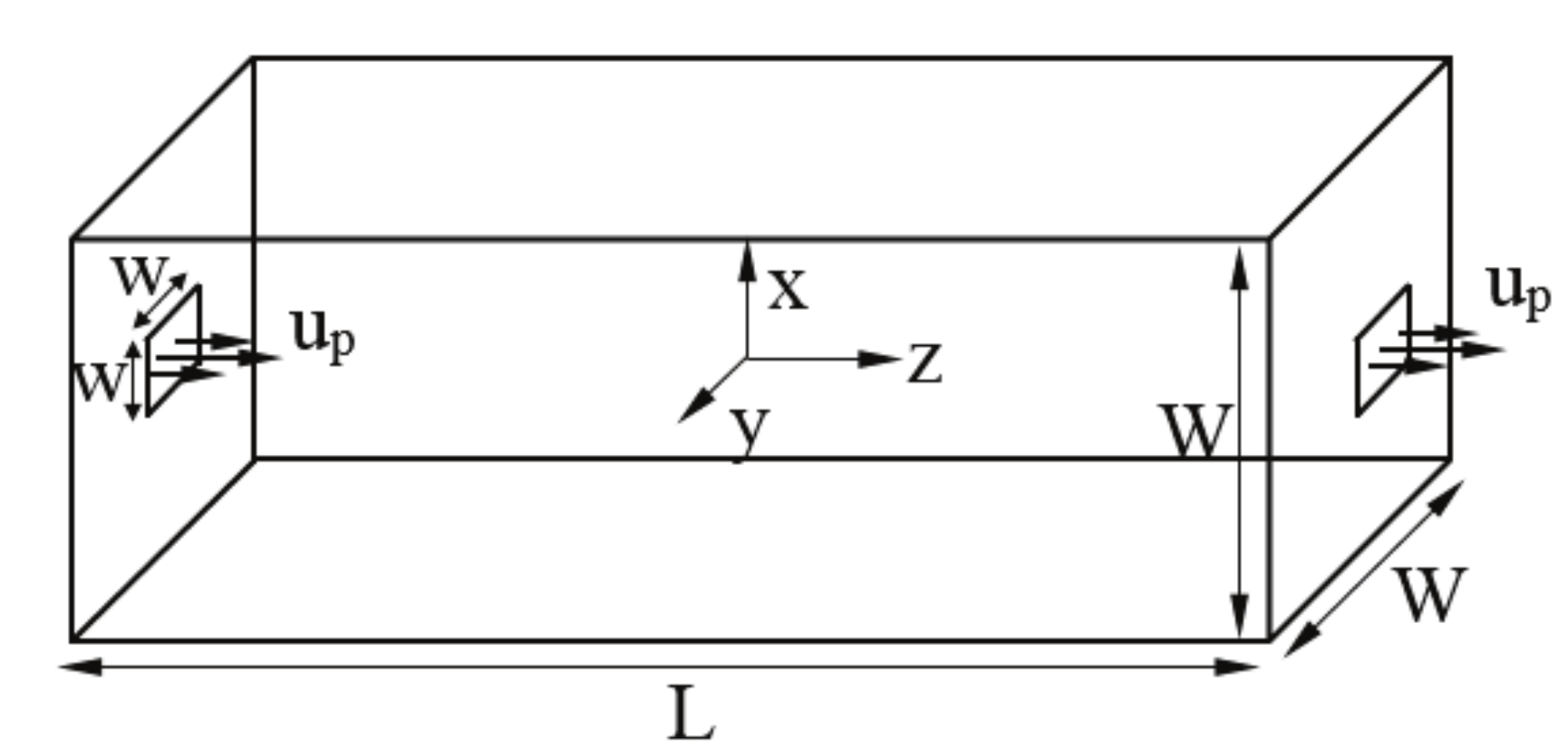}
\caption{Schematic of a balanced inflow-outflow configuration in a 3D channel with a square cross section. The origin of the coordinate system is at the center of the channel. The inflow (and outflow) velocity profiles are given by $u_z(x,y,z=\pm L/2)=u_p[1-(2x/w)^2][1-(2y/w)^2]$ for $|x|,|y|\leq w/2$ and zero otherwise.}
\label{fig:3Dschematic}
\end{figure}
\begin{figure}[!htbp]
\centering
\includegraphics[width=\columnwidth]{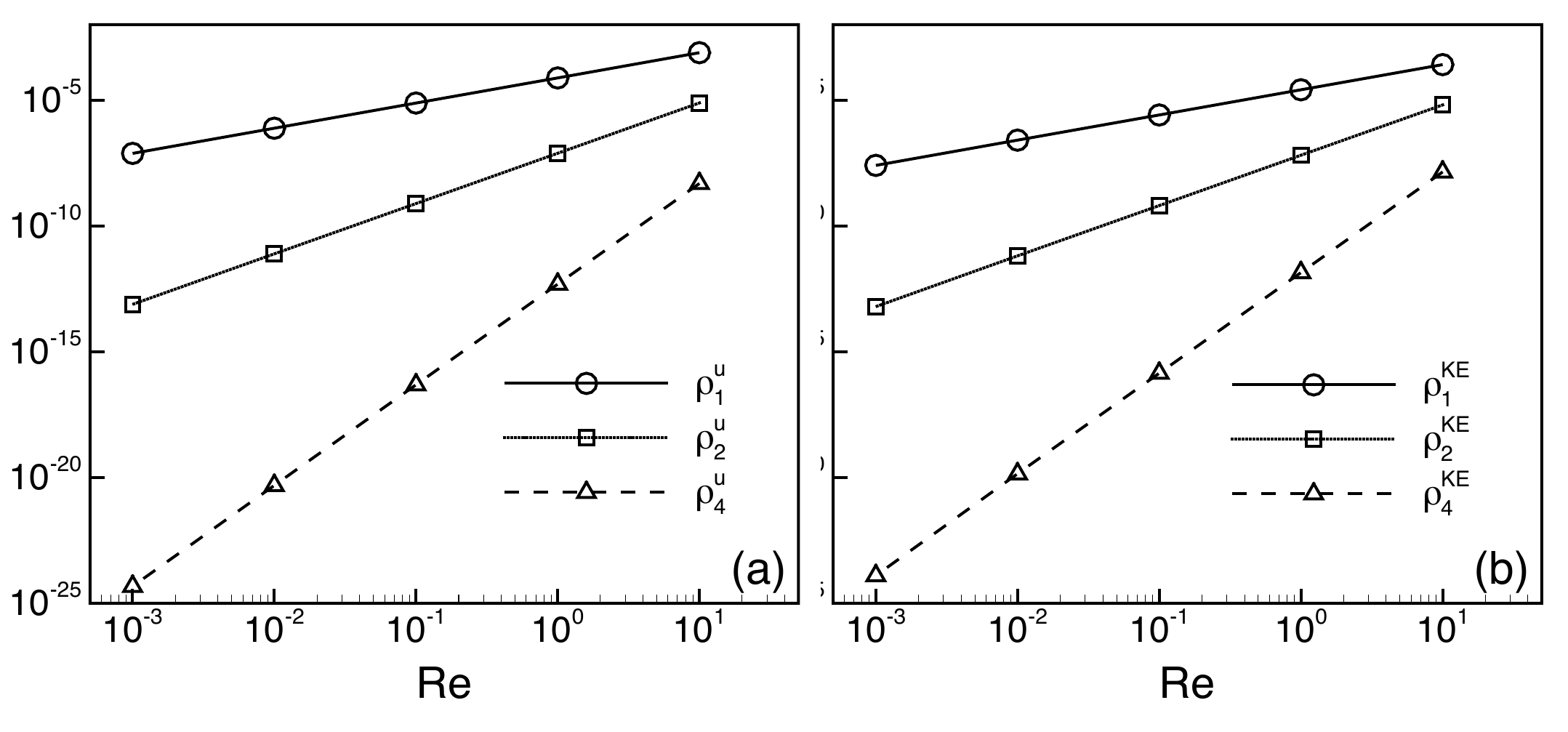}
\caption{Dependence of $\mathcal{PT}$ asymmetries on Reynolds number for laminar, viscous fluid flow in a three-dimensional channel with balanced inflow and outflow. Panel (a) shows that the asymmetries in velocity scale as a power-law and panel (b) shows that the same holds for kinetic energy density asymmetries, $\rho^i_n= A^i_{3D}\mathrm{Re}^n$ ($i=u,\mathrm{KE}$), over four decades in Reynolds number. The range of $\mathrm{Re}$ is chosen to ensure that the three dimensional flow is laminar.}
\label{fig:3DthreerhoRe}
\end{figure}

Figure~\ref{fig:3Dschematic} shows the schematics of a 3D channel with square cross-section of area $W^2$ and length $L$. The origin of the coordinate system is at the center of the channel. The flow inlet is a square of side $w$ located at $z=-L/2$ and the outlet is of the same size located at $z=+L/2$. For the three-dimensional case, the parity operator is given by $\mathcal{P}:{\bf r}\rightarrow -{\bf r}$, and the velocity field is $\mathcal{PT}$ symmetric if $\mathcal{PT}{\bf u}({\bf r})={\bf u}(\mathcal{P}{\bf r})={\bf u}(-{\bf r})$. We impose balanced  $\mathcal{PT}$-symmetric inflow and outflow as follows,
\begin{eqnarray}
\label{eq:uz}
u_z(x,y,z=L/2) & = & u_p\left[1-(2x/w)^2\right]\left[1-(2y/w)^2\right]\nonumber\\
& = & u_z(-x,-y,z=-L/2).
\end{eqnarray}
The 3D MRT-LBM is used to obtain the steady-state velocity field ${\bf u}({\bf r})$. The $\mathcal{PT}$-asymmetry function in the velocity field is defined as
\begin{equation}
\rho^u_n=\frac{1}{2W^2L u_p^n}\int^{L/2}_{-L/2}dz\int^{W/2}_{-W/2}dx\,dy\,
|\mathcal{PT}{\bf u}({\bf r})-{\bf u}(-{\bf r})|^n,
\end{equation}
and an analogous expression defines kinetic energy density asymmetries $\rho^\mathrm{KE}_n$. In contrast to the two-dimensional systems, the vorticity in three dimensions is a pseudovector. Therefore, here, we restrict ourselves only to $\mathcal{PT}$ asymmetries in velocity and kinetic energy density. Figure~\ref{fig:3DthreerhoRe} shows the dependence of $\mathcal{PT}$-asymmetries on the Reynolds number over four decades, $10^{-3}\leq\mathrm{Re}\leq 10$. We emphasize that these low values of Reynolds number are used to ensure a laminar flow in three dimensions. Panel (a) in Fig.~\ref{fig:3DthreerhoRe} shows that the velocity asymmetries have a power-law scaling with Reynolds number, $\rho^u_n\propto\mathrm{Re}^n$, for $n=1,2,4$. Panel (b) shows an identical behavior for the kinetic energy density asymmetry, $\rho^\mathrm{KE}_n\propto\mathrm{Re}^n$. These preliminary results suggest that the power-law scaling in ``balanced inflow-outflow configurations'' is robust and remains valid in three dimensional systems with laminar flow.


\section{Discussion}
\label{se:disc}
In this paper, we have developed the formalism for ``balanced inflow-outflow configurations'' of incompressible viscous flow. We have defined configuration-dependent asymmetries for the steady-state fluid velocity ${\bf u}({\bf r})$ , kinetic energy density $\varrho{\bf u}^2({\bf r})/2$, and vorticity $\omega_z({\bf r})$, and obtained their dependence on the Reynolds number. The nonlinearities due to convective acceleration, the $({\bf u}\cdot\nabla){\bf u}$ term in the Navier-Stokes equation make it difficult to analytically predict the symmetry properties of these observables.

Through numerical simulation via lattice Boltzmann method, we have found that for $\mathcal{PT}$-symmetric configurations {\it all asymmetries $\rho_n$ scale with the Reynolds number with exponent $n$} over a wide range of geometries and boundary velocity profiles in {\it two and three dimensions.} We have also shown that asymmetries in $\mathcal{PT}$-symmetric systems, particularly at low Reynolds numbers $\mathrm{Re}\lesssim 1$, are orders of magnitude smaller than those in systems with traditional, fully-developed-outflow boundary condition.

Our results raise a number of interesting questions, particularly for 3D systems. Does the power-law scaling of asymmetry persist when the flow becomes transitional or turbulent at higher Reynolds numbers? Does the onset of turbulence occur at the same Reynolds number for a balanced geometry as it does for the traditional, fully-developed-outflow geometry? Why does the single-power-law scaling remain valid for $\mathrm{Re}\gg 1$, when it is expected to be valid only for small Reynolds numbers? Is  such robustness a consequence of the $\mathcal{PT}$ symmetric inflow-outflow conditions? Answers to these questions will not only improve the understanding of the pertinent physics of viscous flow, but will also inspire innovative flow control techniques with implications to a wide variety of fields.


\section{Acknowledgment}
\label{sec:ack}
This work was supported by IUPUI Research Support Fund Grant (HY), the  National Nature Science Foundation of China NSFC-11072229 (HY), and the National Science Foundation DMR-1054020 (YJ).


\end{document}